\documentclass[prx,superscriptaddress,reprint]{revtex4-2}
\usepackage{amstext}
\usepackage{amssymb} 
\usepackage{amsmath}
\usepackage{amsthm} 
\usepackage{algorithm}
\usepackage{algpseudocode}
\usepackage{hhline}
\usepackage{dsfont}
\usepackage{float}
\usepackage{tikz}
\usepackage{makecell}
\usepackage{siunitx}
\usepackage{bbold}
\usepackage[braket,qm]{qcircuit}
\usepackage{braket}
\usepackage{mathtools} 
\usepackage{xcolor}
\usepackage{hyperref}
\usepackage{enumitem}
\usepackage{natbib}
\hypersetup{colorlinks=true,allcolors=[rgb]{0.1,0.1,0.5}}

\usepackage{cleveref}
\Crefname{equation}{Eq.}{Eqs.}
\Crefname{figure}{Fig.}{Figs.}
\Crefname{tabular}{Tab.}{Tabs.}

\theoremstyle{remark}

\definecolor{mygreen}{rgb}{0,0.5,0}

\newcommand{\tabincell}[2]{\begin{tabular}{@{}#1@{}}#2\end{tabular}}
\newcommand{\normord}[1]{:\mathrel{#1}:}

\begin{document}

\title{Waveguide QED toolboxes for universal quantum matter}

\author{Y. Dong}
\affiliation{Research Center for Quantum Sensing, Zhejiang Lab, Hangzhou 311121, China}
\affiliation{Institute for Quantum Computing and Department of Physics \& Astronomy, University of Waterloo, Waterloo, Ontario N2L 3G1, Canada}
\author{J. Taylor}
\affiliation{Institute for Quantum Computing and Department of Physics \& Astronomy, University of Waterloo, Waterloo, Ontario N2L 3G1, Canada}
\affiliation{Q-Block Computing Inc., Kitchener, Ontario N2C 2C8, Canada}
\author{Y. S. Lee}
\affiliation{Institute for Quantum Computing and Department of Physics \& Astronomy, University of Waterloo, Waterloo, Ontario N2L 3G1, Canada}
\author{H. R. Kong}
\affiliation{Institute for Quantum Computing and Department of Physics \& Astronomy, University of Waterloo, Waterloo, Ontario N2L 3G1, Canada}
\affiliation{Q-Block Computing Inc., Kitchener, Ontario N2C 2C8, Canada}
\author{K. S. Choi}
\email{kyung.choi@uwaterloo.ca}
\affiliation{Institute for Quantum Computing and Department of Physics \& Astronomy, University of Waterloo, Waterloo, Ontario N2L 3G1, Canada}
\affiliation{Q-Block Computing Inc., Kitchener, Ontario N2C 2C8, Canada}
\affiliation{Perimeter Institute for Theoretical Physics, Waterloo, Ontario N2L 2Y5, Canada}

\begin{abstract}
{\noindent{An exciting frontier in quantum information science is the realization and control of complex quantum many-body systems. Hybrid nanophotonic system with cold atoms has emerged as the paradigmatic platform for realizing long-range spin models from the bottom up, exploiting their modal geometry and group dispersion for tailored interactions. An important challenge is the physical limitation imposed by the photonic bath, constraining the types of local Hamiltonians that decompose the available physical models and restricting the spatial dimensions to that of the dielectric media. However, at the nanoscopic scale, atom-field interaction inherently accompanies significant driven-dissipative quantum forces that may be tamed as a new form of a mediator for controlling the atomic internal states. Here, we formulate a quantum optics toolbox for constructing a universal quantum matter with individual atoms in the vicinity of 1D photonic crystal waveguides. The enabling platform synthesizes analog quantum materials of universal $2$-local Hamiltonian graphs mediated by phononic superfluids of the trapped atoms. We generalize our microscopic theory of analog universal quantum simulator to the development of dynamical gauge fields. In the spirit of gauge theories, we investigate emergent lattice models of arbitrary graphs, for which strongly-coupled SU$(n)$-excitations are driven by an underlying multi-body interaction. As a minimal model in the infrared, we explore the realization of an archetypical strong coupling quantum field theory, SU($n$) Wess-Zumino-Witten model, and discuss a diagnostic tool to map the conformal data of the field theory to the static and dynamical correlators of the fluctuating photons in the guided mode.}}
\end{abstract}

\maketitle
\section{Introduction}

One of the central problems in quantum information science and condensed matter physics is to create and control strongly interacting quantum systems, and to measure the equilibrium and non-equilibrium properties of the many-body system \cite{Lloyd1996,Amico2008, Kimble2008}. Recent experiments with ultracold atoms have extended the ranges of unconventional phenomena that may be accessed. A common thread to these efforts is the quest to design the Hamiltonian by harnessing the natural interactions available between cold atoms \cite{Bloch2008}. Much of the focus has largely been on analog and Floquet quantum systems. However, these approaches are limited in their applicability to complex target Hamiltonians whose description departs significantly from the microscopic model of the simulator. 

A parallel development has been the exploration of computational complexity of local Hamiltonians, whose ground state properties cannot be efficiently obtained even by a digital quantum computer. An example of such a Quantum-Merlin-Arthur (QMA) problem is to find the ground state of  $2$-local Hamiltonians $\hat{H}_{\text{QMA}}=\sum_{ij} \hat{h}_{ij}$, where the local decomposition $ \hat{h}_{ij}$ consists of at most $2$-body SU$(2)$ operators. More generally, arbitrarily complex quantum matter $\hat{H}_{\text{target}}$ can be emulated with a seemingly simpler but QMA-complete lattice model $\hat{H}_{\text{QMA}}$ \cite{Cubitt2018}, in that all physical properties and local structures of $\hat{H}_{\text{target}}$ can be efficiently mapped onto the universal model $\hat{H}_{\text{QMA}}$. Likewise, a quantum simulator that realizes analogue Hamiltonians $\hat{H}_{\text{QMA}}$ can be adapted for universal quantum computation in the spirits of cellular automata and Hamiltonian computation \cite{Biamonte2008,Nagaj2008,Vollbrecht2008}.

With recent developments in atom-photon interfaces with photonic crystals  \cite{John1990,Kurizki1990,John1996,Hung2013,Thompson2013, Tiecke2014, Goban2014, Goban2015,Hood2016,Samutpraphoot2020,Dordevic2021}, there has been significant interest towards assembling quantum many-body systems by garnering the control over individual quantum systems  \cite{Lloyd1996,Amico2008, Bloch2008, Kimble2008}. With the atomic transition frequency residing within the photonic band gap (PBG), the underlying lattice of atoms cannot dissipate propagating waves into the guided modes (GMs) of the photonic structure. However, the mere presence of the atoms at sites $i,j$ in a waveguide seeds dynamic defect modes that support stable atom-field bound states in the form of evanescent waves \cite{John1990,Kurizki1990,Douglas2015,Shi2016,Calajo2016}, mediating exchange interaction $J_{|i-j|}\vec{\hat{\sigma}}^{(i)}\cdot\vec{\hat{\sigma}}^{(j)}$ between the trapped atoms  \cite{Gullans2012, Gonzalez-Tudela2015}. With auxiliary Raman sidebands and digital time-steps  \cite{Lloyd1996}, the phase-amplitude function $J_{|i-j|}$ can be engineered for atoms coupled to 1D and 2D photonic crystal waveguides (PCW), and realize translationally-invariant pairwise models for quantum magnetism, constrained by the dimension of the dielectric \cite{Hung2016}. Conversely, photons propagating through the guided mode exhibit novel quantum transport and many-body phenomena \cite{Hartmann2006,Greentree2006,Ramos2014,Lodahl2017,Pichler2017}.

\begin{figure*}
  \includegraphics[width=2\columnwidth]{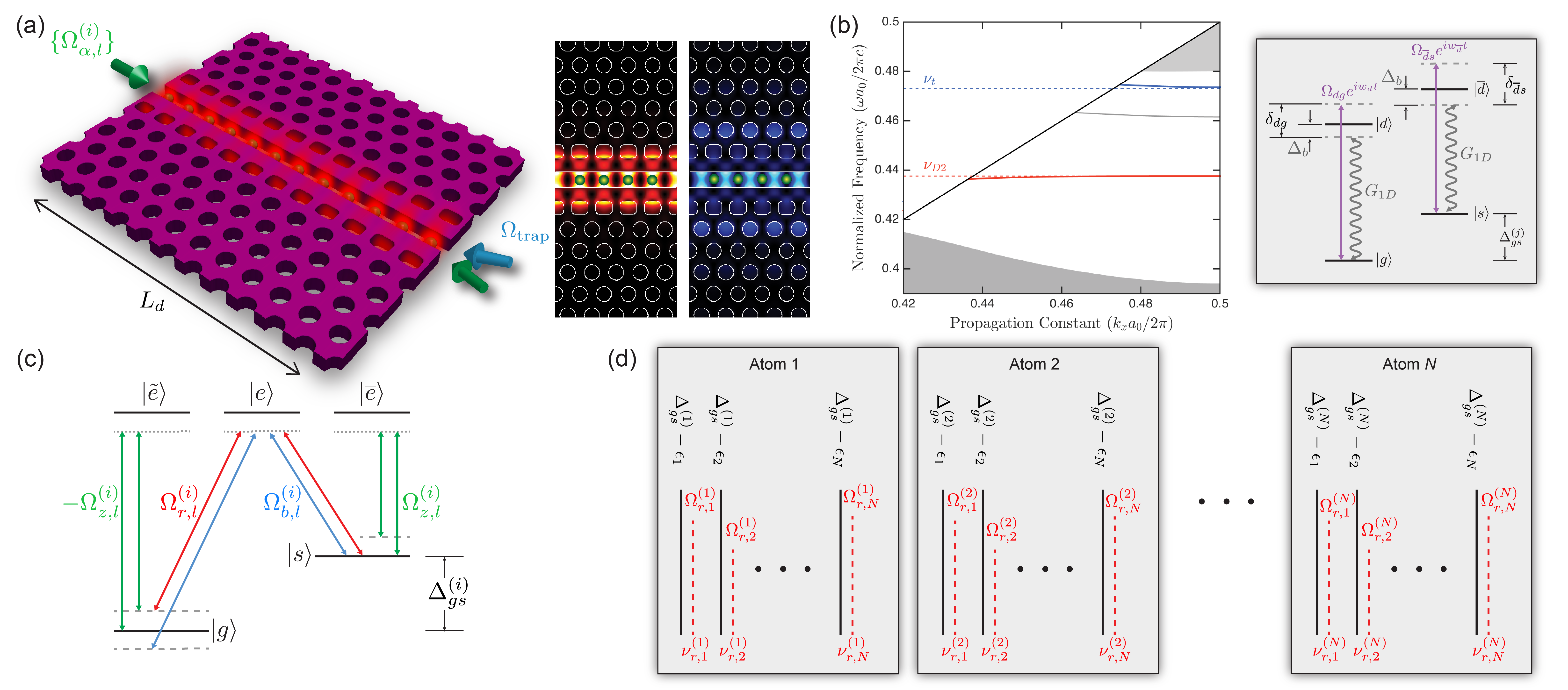}
  \caption{{Complex quantum many-body physics with waveguide QED systems.} (a) Exemplary waveguide QED spin network. Slotted squircle photonic crystal waveguide (SPCW) enables a versatile platform for highly tunable defect guided modes, with the supermodes shown in the inset. As a candidate PCW, structural parameters are provided in the Table \ref{SPCWtable} and discussed in Appendix \ref{Appendix_SPCWsection}. Green spheres represent the trapped atoms. Inset. Contour map of the intensity profile for TE supermodes for exciting (trapping) Cs atoms at wavelengths $\lambda_p=852$ nm ($\lambda_t=794$ nm). (b)  Normalized band diagram for the supermodes of SPCW. Inset. Two lasers $\Omega_{dg},\Omega_{\overline{d}s}$ with detunings $\delta_{dg},\delta_{\overline{d}s}$ create strong photonic Lamb shifts $\sim e^{-|i-j|a_0/L_c}$ between two atoms localized within a photonic bandgap  \cite{Douglas2015}. The bandgap is detuned by $\Delta_b$ with respect to the transition frequency. (c) Raman couplings synthesize programmable interactions between two atoms at sites $i,j\in\{1\cdots N\}$ for any combination of SU($2$) spin operators. Site-resolved addressing with spatially global fields $\Omega_{\alpha,l}^{(i)}$ (in the frequency domain) is achieved through inhomogeneous Zeeman shifts $\Delta_{gs}^{(i)}$ through intermediated excited states $|\tilde{e}\rangle,|e\rangle, |\overline{e}\rangle$ with $\alpha\in\{r,b,z\}$. (d) Raman engineering. Programmable Raman fields $\Omega_{\alpha,l}^{(i)}$ selectively couples internal states $|g\rangle, |s\rangle$ of atom $i$ to the Bogoliubov phononic mode $l\in\{1,\cdots,N\}$ with two-photon detuning $\nu_{l}^{(i)}$. Each single sideband mode with frequency $\nu_{l}^{(i)}$ (red dash line) is nearly resonant to $\Delta_{gs}^{(i)}-\epsilon_l$ (black solid line), where $\epsilon_l$ is the phonon spectrum. Only the red sideband couplings are depicted for simplicity.}
  \label{fig1}
  \end{figure*}

At the nanoscale, atom-field interaction is modified by the electromagnetic vacuum of the ``dielectric,'' consisting of both the passive photonic structure and the active emitters. Such a quantum dielectric is inherently renormalized by the strong coherent and dissipative radiative forces between the atoms. Indeed, complex spin-mechanical textures arise through localized spin-dependent photon-mediated forces  \cite{Manzoni2017}. More generally, nanoscopic quantum forces modify the mechanical ``vacuum" of the atomic motion, where Bogoliubov phonons are distributed across the atomic sample as a collective bath that in turn couples to the spin system. Dissipative nature of these forces in PCW may be exploited to stabilize and self-organize new forms of mechanical phases of quantum matter, and complex observables may be constructed for the detection of highly entangled quantum systems.

Here, we harness the coherent coupling between atomic motion and internal states in 1D PCW for the realization of an analogue universal quantum matter. We develop a low-energy theory for the quantum motion of the trapped atoms in the bandgap regime of waveguide quantum electrodynamics (QED). By coupling Bogoliubov phonons to the spin matter, we realize a fully programmable lattice spin system $\hat{\rho}_s$ for neutral atoms. In our approach, arbitrary binary interaction $\hat{h}_{ij}\simeq \sum_{\alpha,\beta}J^{(i,j)}_{\alpha\beta}\hat{\sigma}_{\alpha}^{(i)}\hat{\sigma}_{\beta}^{(j)}$ is realized for any combination of SU($2$)-spin operators $\hat{\sigma}_{\alpha}^{(i)},\hat{\sigma}_{\beta}^{(j)}$ with $\alpha,\beta\in\{0,x,y,z\}$ between sites $i$ and $j$. Our spin-network $\hat{\rho}_s$ are described by Hamiltonian graphs with connectivity $i,j$ that can no longer be represented by spatial lattices and dimensions, and realizes the universal $2$-local quantum matter $\hat{H}_{\text{QMA}}=\sum_{i,j}\hat{h}_{ij}$ in a fully analog manner. 
 
Moreover, we formulate a hardware-efficient protocol to design dynamical gauge structures of many-body system and realize a plethora of SU($n$) models with our waveguide QED simulator. Motivated by gauge fixing in quantum spin glasses and color codes, we describe a general construction for which the low energy physics of $\hat{\rho}_s$ encompasses the full scope of binary lattice models for SU($n$)-spin excitations with local constraints that protect the many-body wave function $\hat{\rho}_s$ from errors. Here, atomic arrays constrained by their local symmetries are encoded into logical SU($n$)-blocks, and dynamical U($1$)-gauge fields mediate programmable long-range interactions between the logical blocks. 

Utilizing these capabilities, we demonstrate the versatility of our universal analog simulator by constructing chiral spin liquids \cite{Balents2010} and holographic strange metals  \cite{Sachdev2015, Kitaev2015}. As a primordial example to the tower of phases, we explore the physics of SU($3$) Wess-Zumino-Witten conformal field theory (CFT), a holographic dual to a Chern-Simons gravity \cite{Witten1984}, by encoding the target CFT onto the low-energy sector of our waveguide QED simulator. We investigate the critical scaling of CFT entanglement and the dynamics of semionic quasiparticle excitations, as reflected by the fluctuating photons of the PCW. Our networked approach provides powerful tools for controlling analog quantum systems with complexities far beyond of regular spin lattices heretofore explored. 

\section{Platform}

\subsection{Lamb shifts in PCW: Phononic Hubbard model}\label{Lamb_section}

Our approach is based upon the unique capability of PCWs to induce strong photon-mediated forces between proximal neutral atoms and to create many-body states of internal spin and external motion. By engineering the QED vacuum of the PCW, we synthesize coherent mechanical coupling between the trapped atoms, and renormalize the atomic array into a mechanical quantum network. Long-range interaction of the universal Hamiltonian $\hat{H}_{\text{QMA}}=\sum_{i,j}\hat{h}_{i,j}$ is mediated through the phononic quantum channels with full control over the decompositions $\hat{h}_{i,j}$ and their connectivity $i,j$.

As shown in Fig. \ref{fig1}, our basic building block is an 1D lattice of neutral atoms at positions $x_i$ strongly coupled to a dispersive PCW with mode function $u_{k_0}(x)$ represented by the red line of Fig. \ref{fig1}(b). The band edge at frequency $w_b$ is red-detuned by $\Delta_b=\omega-\omega_b>0$, so that the atomic transition frequency $\omega$ lies within the band gap. Each atom is tightly localized at the antinodes of $u_{k_0}(x)$ with trap frequency $\omega_t$ and lattice constant $a_0$ by a nanoscopic optical potential ${V}_{\text{T}}=V_0\sin^2 k_0 x$ with a trapping field at a higher-order GM (blue line). In Appendix \ref{Appendix_SPCWsection}, we analyze a versatile candidate structure (Silicon Nitride Squircle PCW) with highly tunable GMs in terms of TE photonic band gap, effective photon mass $m_e$ and mode area $A_{\text{eff}}$ near the band edge $k_x=k_0$ (See Fig. \ref{fig1}(b) for the band diagram). $|g\rangle$ and $|s\rangle$ are the two hyperfine ground states that define the computational basis $\mathcal{C}$ of the waveguide QED simulator, and the ground states respectively couple to excited states $|d\rangle$ and $|\overline{d}\rangle$, which will be eliminated to induce a pure mechanical coupling between the atoms.

The atom-PCW Hamiltonian reads $\hat{H}_\text{PCW} = \int d \boldsymbol{x}\int_0^\infty d\omega \hat{\boldsymbol{f}}^\dagger(\boldsymbol{x},\omega)\hat{\boldsymbol{f}}(\boldsymbol{x},\omega) + \sum_i ( \omega_d\sigma_{dd}^{(i)}+\omega_{\bar{d}} \sigma_{\bar{d}\bar{d}}^{(i)}+\Delta_{gs}\sigma_{ss}^{(i)} ) +\sum_{i=1}^{N_a}\sum_{\mu=dg,\bar{d}s}[\int_0^\infty d\omega \boldsymbol{E}(\boldsymbol{x}_i,\omega)\cdot\boldsymbol{d}_\mu\hat{\sigma}_{\mu}^{(i)}+  \Omega_{\mu}\sigma_{\mu}^{(i)}e^{-i\nu_{\mu}t}]$, where $\boldsymbol{d}_{dg(\bar{d}s)}$ is the transition dipole momentum from $|g(s)\rangle$ to $|d(\bar{d})\rangle$ and $\Omega_{dg(\bar{d}s)}$ is the Rabi frequency of the pumping fields with frequency $\nu_{dg(\bar{d}s)}$ that couples  $|g(s)\rangle$ and $|d(\bar{d})\rangle$. We assume $\boldsymbol{d}_{dg}=\boldsymbol{d}_{\bar{d}s}=\boldsymbol{d}$. The electric field in the PCW can be represented by classical Green's function $\boldsymbol{G}(\boldsymbol{x},\boldsymbol{x}^\prime,\omega)$ as $\hat{\boldsymbol{E}}(\boldsymbol{x},\omega)=i\mu_0\omega^2\sqrt{\frac{\epsilon_0}{\pi}}\int d\boldsymbol{x}^\prime \sqrt{\text{Im}\{\epsilon(\boldsymbol{x}^\prime,\omega)\}}\boldsymbol{G}(\boldsymbol{x},\boldsymbol{x}^\prime,\omega)\hat{\boldsymbol{f}}(\boldsymbol{x},\omega)$, where $\hat{\boldsymbol{f}}(\boldsymbol{x},\omega)$ represents the quantized excitation of the dielectric with permittivity $\epsilon(\boldsymbol{x}^\prime,\omega)$ \cite{Gruner1996}.

In the limit $f_{dg(\bar{d}s)}=\Omega_{dg(\bar{d}s)}/\delta_{dg(\bar{d}s)}=f\ll 1$ where $\delta_{dg(\bar{d}s)} = \nu_{dg(\bar{d}s)} - \omega_{d(\bar{d})}$, we adiabatically eliminate the excited states $|d\rangle$ and $|\bar{d}\rangle$ from the system and integrate out the photonic modes \cite{Dung2002, Hughesstudentthesis, Asenjo-Garcia2017, Asenjo-Garcia2017b}. We thereby obtain the low-energy Liouvillian dynamics $\dot{\hat{\rho}}=-i[\hat{H}^{\text{int}}_M,\hat{\rho}]+\mathcal{L}_0[\hat{\rho}]+\mathcal{L}_M[\hat{\rho}]$ with a state-independent Hamiltonian 
\begin{equation}
  \hat{H}^{\text{int}}_M = f^2 \Delta_{\text{Lamb}}(\hat{\boldsymbol{x}}_i,\hat{\boldsymbol{x}}_j)\hat{\sigma}_0^{(i)}\hat{\sigma}_0^{(j)}
\end{equation} 
and the respective Lindblad superoperators $\mathcal{L}_0[\hat{\rho}] = \sum_{i,j}\frac{\Gamma_{ij}f^2}{2}(2\hat{\sigma}_0^{(i)}\hat{\rho}\hat{\sigma}_0^{(j)}-\hat{\sigma}_0^{(i)}\hat{\sigma}_0^{(j)}\hat{\rho} - \hat{\rho}\hat{\sigma}_0^{(i)}\hat{\sigma}_0^{(j)})$ and ${\mathcal{L}_M}[\hat{\rho}]=\sum_{i,j}\frac{\Gamma_{ij}f^2}{2}(2e^{i k \hat{\boldsymbol{x}}_i}\hat{\rho}e^{-i k \hat{\boldsymbol{x}}_j}-e^{i k (\hat{\boldsymbol{x}}_i-\hat{\boldsymbol{x}}_j)}\hat{\rho} - \hat{\rho}e^{i k (\hat{\boldsymbol{x}}_j-\hat{\boldsymbol{x}}_i)})$ acting on the internal and external degrees of freedom (DOF), where $\hat{\sigma}_0=|g\rangle \langle g|+|s\rangle\langle s|$ is the identity spin operator in  $\mathcal{C}$. Photonic Lamb shift and correlated dissipation, modified by the PCW, are given by
\begin{eqnarray}
	\Delta_{\text{Lamb}}(\hat{\boldsymbol{x}}_i,\hat{\boldsymbol{x}}_j)&=&2 \mu_0\omega_b^2\boldsymbol{d}^\ast\cdot \text{Re}[\boldsymbol{G}_s(\hat{\boldsymbol{x}}_i,\hat{\boldsymbol{x}}_j,\omega_b)]\cdot \boldsymbol{d}\\
	\Gamma_{ij}(\hat{\boldsymbol{x}}_i,\hat{\boldsymbol{x}}_j)&=&\mu_0\omega_b^2\boldsymbol{d}^\ast\cdot \text{Im}[\boldsymbol{G}(\hat{\boldsymbol{x}}_i,\hat{\boldsymbol{x}}_j,\omega_b)]\cdot \boldsymbol{d},
\end{eqnarray}
where $\boldsymbol{G}_s=\boldsymbol{G}-\boldsymbol{G}_0$ is the scattering Green's function relative to the vacuum term $\boldsymbol{G}_0$ (See Appendix \ref{Appendix_SPCWsection}). Seen from the atoms, the correlated radiative decay $\Gamma_{ij}$ does not directly contribute to the dynamics of computational subspace $\mathcal{C}$ but induces mechanical damping to the atomic quantum motion.

In our case, the GM near the band edge $k_x=k_0$ (corresponding frequency $\omega_b$) exhibits an extremely flat band $w_k-w_b\simeq -\frac{1}{2m_e} (k_x-k_0)^2$ and the GM photons acquire large mass $1/m_e=-(\partial^2 w_k/\partial k_x^2)$ (See Fig. \ref{fig1}(b) for the first Brillouin zone). In the reactive regime of the PBG, the atoms predominantly couple to this band edge and the Green's function is approximated by
\begin{equation}
	G_\text{1D}(\hat{{x}}_i,\hat{{x}}_j)=J_{1D}u_{k_0}({x}_i)u_{k_0}({x}_j)e^{-|\hat{{x}}_i-\hat{{x}}_j|/L_c},
\end{equation}
where the localization length $L_c = \sqrt{1/2m_e\Delta_{e}}\sim a_0$ is controlled by the detuning $\Delta_{e}\simeq 2\Delta_{b}$ of the pumping field from band edge. $J_\text{1D}=-\frac{c^2}{2\omega_bL_cA_\text{eff}}\frac{1}{\Delta_{e}+i\kappa/2}$  is the coupling rate to the PCW with effective mode area $A_{\text{eff}}\simeq \lambda^2$, mode function $u_{k_0}({x}_i)$ at the band edge, and decay rate $\kappa$ ($\kappa_0$) in the band gap (at the band edge). The correlated Lamb shift thereby provides the tunnelling interaction $\hat{H}_M=f^2\Delta_{\text{1D}}e^{-|\hat{{x}}_i-\hat{{x}}_j|/L_c}$ between local phonons pinned on the lattice sites ${x}_i$ and ${x}_j$, with $\Delta_{\text{1D}}=\frac{\omega_b d^2\Delta_{e} }{\epsilon_0L_cA_\text{eff}(\Delta_{e}^2+\kappa^2/4)}$. Importantly, the collective damping $\Gamma_{\text{1D}}\sim \tilde{\Gamma}_{\text{1D}}\exp(-L_d/L_c)$ is exponentially inhibited for a finite device length $L_d$, with $\tilde{\Gamma}_{\text{1D}}=\frac{\omega_b d^2\kappa_0}{2\epsilon_0L_cA_\text{eff}(\Delta_{e}^2+\kappa_0^2/4)}$, and figure of merit $\mathcal{F}=\Delta_{\text{1D}}/\Gamma_{\text{1D}}\sim \exp(L_d/L_c)\gg 1$ is favorable for massive photons with flat bands and long device length $L_d\gg L_c$.

For nearest-neighbour coupling $L_c\sim a_0$, we expand the mechanical Hamiltonian $\hat{H}_M$ around the equilibrium positions to the second order of the zero-point motion $x_0=\sqrt{\hbar/2m\omega_t}$ and obtain the quadratic mechanical Hamiltonian
\begin{equation}\label{mechHam}
	H_\text{M}=\sum_{i}\frac{{\hat{p}_{i}^{2}}}{2m}+\frac{{m\omega_t^{2}}}{2}\hat{x}_{i}^{2}-\frac{\hbar g_\text{m}}{L_c^{2}}\hat{x}_{i}\hat{x}_{i+1}+\mathcal{O}(\hat{x}_i^4),
\end{equation}
for the mechanical coupling constant $g_\text{m}=f^2\Delta_\text{Lamb}$ and the trap frequency $\omega_t$. With a first-type sine transform $B_{jk}=\frac{1}{\sqrt{N}}\sin[i\frac{\pi}{N+1}jk]$, we diagonalize $H_\text{M}=\sum_{l=1}^{N} \epsilon_{l}\hat{\beta}_l^\dagger \hat{\beta}_l$, whose quasiparticles are the Bogoliubov phonons $\{\hat{\beta}_l\}$ with spectrum $\epsilon_{l}=\sqrt{\omega_{t}^{2}+\frac{2\hbar g_\text{m}x_0^2}{L_c^{2}}\cos(\frac{\pi}{N+1}l)}$ with momentum-space mode indices $l$. For nanoscopic optical potentials with $k_x x_0\not\ll 1$, Eq. \ref{mechHam} describes a 1D Bose-Hubbard model for atomic motion with on-site (long-range) interaction $\sim \hat{x}_i^4/x_0^4$ ($|\hat{x}_i-\hat{x}_j|^4/L_c^4$), where the Bogoliubov phonons are excited out of the superfluid vacuum. The radiative damping $\Gamma_{1D}$ gives rise to motional decoherence ${\mathcal{L}_M}[\hat{\rho}_\text{MA}]=\sum_{l}\frac{\gamma_m}{2}\left(2\hat{\tilde{x}}_{l}\hat{\rho}_\text{MA}\hat{\tilde{x}}_l-\{\hat{\tilde{x}}_{l}^2,\hat{\rho}_\text{MA}\}\right)$
with damping $\gamma_m = \frac{g_m x_0^2}{\Delta_e L_c^2}\kappa e^{-L_d/L_c}$ and quadrature $\hat{\tilde{x}}_l=\hat{\beta}_l+\hat{\beta}_l^{\dagger}$. As further discussed below, the momentum-space Bogoliubov modes constitute the frequency-selective channels of an all-to-all connected mechanical quantum network and coherently mediate the interactions between the atomic nodes, transforming the atomic array into a universal quantum matter.

\subsection{Networked universal quantum matter}\label{universalmodel_section}  

To mediate the universal lattice model via the phononic channels, we gain independent control over the interaction coefficients between any atom pair $i,j$ by way of Raman engineering in the sideband-resolved limit. This is ensured in the reactive regime of PCW, because the mechanical damping constant $\gamma_m$ is exponentially suppressed by $\mathcal{F}\sim  \exp(L_d/L_c)\gg 1$ relative to the phonon spread. As shown in Fig. \ref{fig1}(c), we distinguish the coupling of an individual atom $i$ to a particular Bogoliubov mode $l$ with site-dependent ground-state energy shift $\hat{H}_A=\sum_i\Delta_{gs}^{(i)}\hat{\sigma}_z^{(i)}$ with $\Delta_{gs}^{(i)}=\Delta_{gs}+g_F m_F B(x_i)$ in the form of a linear Zeeman gradient  $B(x_i)$  \cite{Hung2016}. The ground-state shift $\delta\Delta_{gs}$ between neighboring sites is larger than the width of the phonon spectrum $|\epsilon_N-\epsilon_1|$, so that the frequency difference $\Delta_{gs}^{(i)}-\epsilon_l$ is different for all pairs of $(i, l)$ (See Fig. \ref{fig1}(d)). 

Then, we introduce spatially global Raman interaction $\hat{H}=\hat{H}_M+\hat{H}_A+\sum_{i,j}\sum_{\alpha,l}\frac{\Omega^{(j)}_{\alpha,l}}{2}\hat{\sigma}_{\alpha}^{(i)}\sin(k^{(j)}_{{\alpha},l}\hat{x}_{i})e^{-i\nu^{(j)}_{{\alpha},l}t}+h.c$ \cite{Jaksch2003,Bermudez2011, Korenbilt2012,Aidelsburger2013} with $N^2$ frequency sidebands to the atom chain through the GM, where $k^{(i)}_{\alpha,l}\simeq k$ and $\nu^{(i)}_{\alpha,l}$ denote the wavenumber and frequency for the Raman fields that couple the spin operator $\hat{\sigma}_{\alpha}^{(i)}$ of atom $i$ with $\alpha\in\{\pm,z\}$ to the Bogoliubov mode $l$ with $\alpha\in\{\pm,z\}$. By expanding $\sin(k\hat{x}_{i})\simeq\sum_{l}\eta_{0}B_{il}(\hat{\beta}_{l}^{\dagger}+\hat{\beta}_{l})$ in the Lamb-Dicke limit with $\eta_0=x_0/a_0$ and switching to the interaction picture, we find $\hat{H}_\text{MA}=\sum_{\alpha,i,j,l}\Omega^{(j)}_{\alpha,l}\hat{\sigma}_{\alpha}^{(i)}e^{-i(\nu^{(j)}_{\alpha,l}-\zeta_\alpha\Delta_{gs}^{(i)})t} \eta_{0}B_{il}\hat{\beta}_{l}^{\dagger}e^{i\epsilon_{l}t}+h.c$, with $\zeta_\alpha=\pm1,0$ for $\alpha=\pm, z$. As $\Delta^{(i)}_{\alpha,l}=\nu^{(i)}_{\alpha,l}-\zeta_\alpha\Delta_{gs}^{(i)}+\epsilon_{l}\ll |\epsilon_{l}-\epsilon_{l-1}|\ll\delta\Delta_{gs}$, we integrate over the rapidly oscillating terms and leave only the slowly-varying terms $\sim \exp[{i(\nu^{(i)}_{\alpha,l}-\zeta_\alpha\omega_{A}^{(i)}+\omega_{l})t}]$ and obtain the spin-mechanical Hamiltonian $\hat{H}_\text{MA}=\sum_{i,l}\sum_{\alpha\in\{x,y,z\}}\frac{\eta_{0}\Omega^{(i)}_{\alpha,l}}{2}B_{il}\hat{\sigma}_{\alpha}^{(i)}\hat{\beta}_{l}e^{-i\Delta_{l}t}+h.c$, thereby coupling the spin operator $\hat{\sigma}^{(i)}_{\alpha}$ at site $x_i$ to a particular Bogoliubov mode $l$. Here, the detuning $\Delta^{(i)}_{\alpha,l}= \Delta_M$ is chosen to be identical for all phononic mode $l$, atom $i$ and spin operator type $\alpha$, and the Rabi frequencies are transformed as $\Omega^{(i)}_{x,l}=(\Omega^{(i)}_{+,l}+\Omega^{(i)}_{-,l})/2$ and $\Omega^{(i)}_{y,l}=i(\Omega^{(i)}_{+,l}-\Omega^{(i)}_{-,l})/2$. 

By projecting the master equation to the computational subspace $\mathcal{C}$ \cite{Reiter2012}, we obtain the open-system dynamics $\dot{\hat{\rho}}_A = -i[\hat{H}_\text{QMA},\hat{\rho}_A]+\sum_\alpha\mathcal{L}[\hat{\rho}_A]$ for the spin system, governed by the universal Hamiltonian
\begin{equation}
	\hat{H}_{\text{QMA}}\simeq \sum_{i,j,\alpha,\beta}J^{(i,j)}_{\alpha,\beta}\hat{\sigma}_{\alpha}^{(i)}\hat{\sigma}_{\beta}^{(j)}+\sum_{i,\gamma} h_{\gamma}^{(i)} \hat{\sigma}_{\gamma}^{(i)},  \label{twobodyHam}
\end{equation} 
and the correlated dissipation
\begin{equation}
{\mathcal{L}}[\hat{\rho}_A]=\sum_{i,j,\alpha,\beta}\frac{\gamma^{(i,j)}_{\alpha,\beta}}{2}\left(2\hat{\sigma}_{\beta}^{(i)}\hat{\rho}_A\hat{\sigma}_{\alpha}^{(j)}-\{\hat{\sigma}_{\alpha}^{(i)}\hat{\sigma}_{\beta}^{(j)},\hat{\rho}_A\} \right)\label{twobodyHamLindblad}
\end{equation}
for any combination of $\alpha,\beta,\gamma\in\{x,y,z\} $ and between any two spins at sites $i,j$. Importantly, the exchange interaction $J^{(i,j)}_{\alpha,\beta}$ and the bias field $h_{\gamma}^{(i)}$ can be arbitrarily designed by solving a set of nonlinear equations $J^{(i,j)}_{\alpha, \beta}=2\text{Re}[\sum_{l}{\tilde{\Omega}^{(i)}_{\alpha,l}\tilde{\Omega}^{(j)\ast}_{\beta,l}}/{\Delta_M}]$ and $h_{\gamma}^{(i)}=-2\epsilon_{\alpha\beta\gamma}\text{Im}[\sum_{l}{\tilde{\Omega}^{(i)}_{\alpha,l}\tilde{\Omega}^{(i)\ast}_{\beta,l}}/{\Delta_M}]$ where $\tilde{\Omega}^{(i)}_{\alpha,l}=\eta_{0}\Omega^{(i)}_{\alpha,l}B_{il}$ and Levi-Civita symbol $\epsilon_{\alpha\beta\gamma}$. Namely, we have $6N^2$ DOFs for the sidebands $\{\Omega_{\alpha,l}^{(i)}\}$ from the nonlinear equations, while only $3(3N^2-N)/2$ independent parameters $\{J^{(i,j)}_{\alpha, \beta}, h^{(i)}_{\gamma}\}$ are required to represent the universal model $\hat{H}_{\text{QMA}}$. Hence, for any set $\{J^{(i,j)}_{\alpha, \beta}, h^{(i)}_{\gamma}\}$, at least one solution $\{\Omega^{(i)}_{\alpha,l},\Delta_{M}\}$ can be obtained to the target model within certain physical constraints (e.g., laser power). We envisage that the Raman sideband matrices $\{\Omega_{\alpha,l}^{(i)}\}$ are real-time tunable. The Hamiltonian $\hat{H}_{\text{QMA}}(t)$ can be evolved to map out complex phase diagrams of many-body models and be globally quenched to study out-of-equilibrium dynamics. The frequency sidebands $\{\Omega_{\alpha,l}^{(i)}\}$ can be streamed by the time-domain response function $\Omega (t)$ using a single-mode phase-amplitude modulator. Dissipative rate is evaluated as $\gamma_{\alpha,\beta}^{(i,j)}=\frac{\gamma_m}{\Delta_M}J^{(i,j)}_{\alpha,\beta}+\gamma_A\delta_{i,j}$, where $\delta_{ij}$ denotes the Kronecker symbol. Coherence-to-dissipation ratio $C=J_{\alpha\beta}/\gamma_{\alpha\beta}=\mathcal{F}/N \sim  \exp(L_d/L_c)/N\gg 1$ of our simulator improves exponentially in the reactive regime. In practice, $C$ is constrained by $\gamma_A$ due to the finite $\Omega_{\alpha,l}$ and $\Delta_M$ of the Raman fields.

\section{Chiral spin liquids in Kagome lattice}\label{CSL_section}

Frustration in lattice spin systems, in which local energy constraints cannot all be satisfied, can lead to deconfined phases of quantum spin liquids (QSL). In a QSL, quantum fluctuations drive the collective state of the spins into highly entangled quantum matter, such as the resonating-valence bond state in Z$_2$-spin liquids, whose emergent topological properties can only be described in terms of long-range entanglement  \cite{Balents2010}. Unlike gapped Z$_2$-spin liquids, chiral spin liquids (CSL) spontaneously break the time-reversal and parity symmetry, while preserving other symmetries, and host fractional quasiparticle excitations with topological order  \cite{Kalmeyer1987}. Such a CSL is thought to be a parent state of the illusive anyonic superconductor.

\begin{figure}[th!]
\includegraphics[width=1\columnwidth]{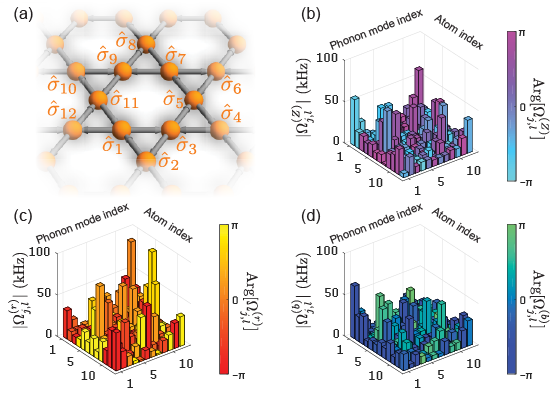}
\caption{{Chiral spin liquid phase in Kagome lattice with vector-spin coupling.} (a)  Antiferromagnetic Heisenberg model $\hat{H}_{\text{AF}}$ with Dzyaloshinskii-Moriya interaction $\hat{\chi}_{\text{vector}}$ is illustrated for spins in an artificial Kagome lattice. The grey arrows indicate the sign of the vector coupling in $\hat{\chi}_{\text{vector}}$. Panels (b)--(d)  Raman sidebands realize $\hat{H}_{\text{CSL}}$ in Eq. \ref{chiralHam} with tunable chirality $\hat{\chi}_{\text{vector}}$ for $J_{\perp}=J_{ZZ}=0.5\text{kHz}$ and $\lambda=0.1\text{kHz}$. Adiabatic evolution through a paramagnetic phase with time-dependent sidebands prepares the chiral spin liquid for cold atoms in PCWs.}\label{CSLfig}
\end{figure}

As an example of Eq. \ref{twobodyHam}, we discuss a method of creating topological CSL discovered by Kalmeyer and Laughlin, a bosonic analogue of the celebrated fractional quantum Hall effect  \cite{Bauer2014,Kumar2015,Essafi2015}, with our waveguide QED toolboxes. We consider an anisotropic antiferromagnetic XXZ Hamiltonian
\begin{equation}
\hat{H}_{\text{CSL}}=\sum_{\langle i j\rangle}\left (J_{\perp}\hat{\sigma}_{\perp}^{(i)}\hat{\sigma}_{\perp}^{(j)}+J_{ZZ}\hat{\sigma}_z^{(i)}\hat{\sigma}_z^{(j)}\right )+\lambda \hat{\chi},\label{chiralHam}
\end{equation}
on a Kagome lattice with tunable spin-chirality $\hat{\chi}$. Despite the physical dimension of the atomic lattice in 1D PCWs, our toolboxes allow the spins to sit on a synthetic geometry provided by the connectivity of the translationally-variant spin-exchange couplings, as depicted by the 2D Kagome lattice in Fig. \ref{CSLfig}. With $\lambda=0$, $\hat{H}_{\text{CSL}}$ reduces to the Kagome XXZ antiferromagnet, which has been widely studied for its time-reversal symmetric Z$_2$ spin liquid  \cite{Balents2002,Isakov2011}.

In the presence of strong chiral interactions on the triangles $\Delta$ of the sublattice, e.g., scalar spin-chirality $\hat{\chi}_{\text{scalar}}=\sum_{i,j,k\in \Delta}\vec{\hat{\sigma}}_i\cdot (\vec{\hat{\sigma}}_j\times\vec{\hat{\sigma}}_k)$, the ground state supports a topologically-protected chiral edge mode circulating the macroscopic outer boundary with closed loops within the inner hexagons of the Kagome lattice  \cite{Bauer2014}. As a convention, the sum $\sum_{i,j,k\in \Delta}$ runs clockwise over the nearest-neighbor sites around the triangles. To see how the extended chiral edge modes emerge in a Kagome lattice, we first identify that the ground state of a single closed loop $\hat{\chi}_{\text{scalar}}$ around a single triangle is the Kalmeyer-Laughlin wavefunction. By mapping the elementary triangular puddles into a Kondo-type network for edge states  \cite{Bauer2014}, individual puddles encircled with the chiral states merge together to develop a macroscopic puddle with a single chiral topological edge state around the outer boundary of the lattice, reminiscent of the two-channel Kondo problem. This allows for unidirectional spin transport along the boundary, and the bulk excitations are described by semionic exchange statistics ($\phi=\pi$).

The difficulty in realizing Eq. \ref{chiralHam} as the low-energy theory of physical Hamiltonians with cold atoms is the spin-chiral coupling $ \vec{\hat{\sigma}}_j\times\vec{\hat{\sigma}}_k$ that breaks the parity symmetry. The capability to realize universal pairwise interaction, including off-diagonal spin operators $\hat{\sigma}_{\alpha}\hat{\sigma}_{\beta}$, makes our approach highly suitable for analog quantum simulation of quantum liquids with chiral spin coupling. As an example, we realize here the minimal instance of CSL with $2$-body vector chirality $\hat{\chi}_{\text{vector}}=\sum_{i,j\in \Delta}\hat{z}\cdot(\vec{\hat{\sigma}}^{(i)}\times \vec{\hat{\sigma}}^{(j)})$ in the form of Dzyaloshinskii-Moriya (DM) interaction. The Raman sideband matrices shown in Figs. \ref{CSLfig}(b)--(d) realize Eq. \ref{chiralHam} on a unit cell of a Kagome lattice in Fig. \ref{CSLfig}(a). The DM interaction breaks the underlying SU$(2)$ symmetry, while preserving the lattice and U$(1)$ spin symmetry. Hence, unlike the case of $\hat{\chi}_{\text{scalar}}$, the CSL does not persist for $\hat{\chi}_{\text{vector}}$ in the limit of strong coupling $\lambda\gg J_{\perp}=J_{ZZ}$. However, it is numerically predicted that gapped CSL phase does exist for XXZ antiferromagnets with a finite vector spin-chirality $\lambda<J_{\perp}=J_{ZZ}$ at zero magnetic field  \cite{Kumar2015, Essafi2015}. The capability to tune vector-chirality as well as other spin-orbit couplings also opens the route to synthetic multiferroics and emergent interfacial spin textures, including skyrmions and topological surface states.

\section{Gauging waveguide QED simulator to interacting SU($n$) lattice models}\label{gaugedSUN_section}

The native Hamiltonian of our waveguide QED simulator spans the universal binary analog models  of SU($2$)-spin operators. In analogy to lattice gauge theories (LGT) that give rise to constrained Hilbert space  \cite{Gingras2014,Castelnovo2008}, we can also design dynamical gauge structures that mediate a wide range of binary models consisting of SU($n$)-operators in a completely analog fashion, such as the Heisenberg quantum magnet for interacting SU($n$)-spins. While the universal quantum simulator can emulate the dynamics of arbitrary unitary dynamics, we confine our discussion here to binary SU($n$)-spin models that arise within the projected gauge-invariant subspace of the parent's SU($2$) waveguide QED simulator. Such a ``condensed matter'' approach  \cite{Gingras2014} can create deconfined quantum phase by direct cooling to its ground state, and the errors can be mitigated within the gauge sector of interest. In this section, we discuss a general Heisenberg SU($n$) quantum magnet as an exemplary implementation, but more complex models involving vector and anisotropy can be realized in an analogous fashion.

\begin{figure}[t!]
\includegraphics[width=1\columnwidth]{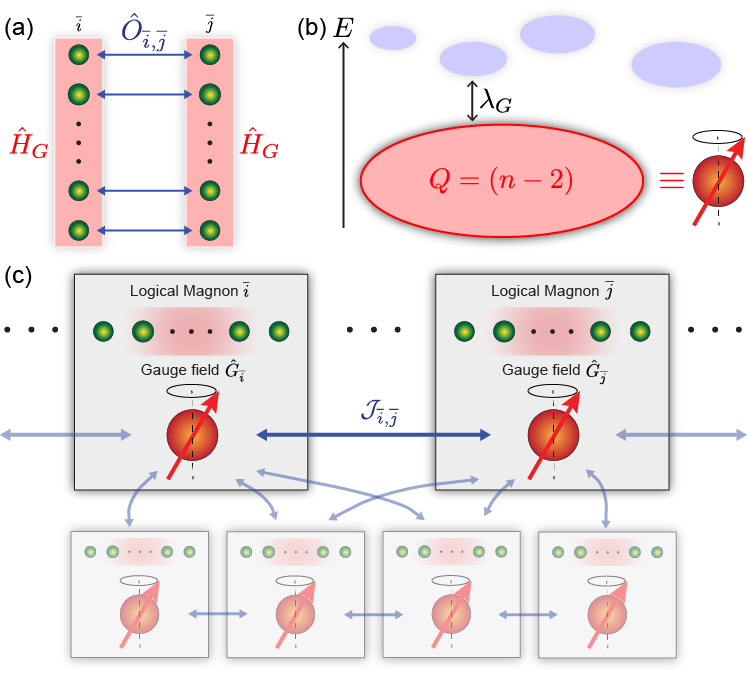}
\caption{{SU($n$)-spin networks under spin ice gauge constraints.} (a)  Parent spin ice Hamiltonian. Trapped atoms in PCWs are subjected to local ``ice"  rules (Gauss laws) with an energetic cost $\hat{H}_G=\lambda_G \sum_{\overline{i}}\hat{G}_{\overline{i}}^2$ within logical blocks $\overline{i},\overline{j}$. Quantum dynamics among the ice states is induced by a perturbative spin-exchange $\hat{O}_{\overline{i},\overline{j}}$ between atoms belonging to different blocks. (b)  Effective reduction of the Hilbert space into gauge sectors. The low-energy dynamics is constrained within the SU($n$) single-excitation sector, represented by a gauge charge $Q=n-2$, with errors protected by a many-body gap $\lambda_G$. (c) The global spin network is transformed into a network of logical SU($n$) spins $\overline{i},\overline{j}$ by encoding the SU($n$)-spin with a collection of $n$ SU($2$)-spins. U$(1)$-gauge constraints $\hat{G}_{\overline{i}}$ block the excitation manifold within the logical spin so that the energy sectors of the parent Hamiltonian are separated by the total excitation number. Spin-exchange coupling between atoms belonging to different logical blocks $\overline{i},\overline{j}$ induces an effective two-body interactions between SU($n$) spins.}
\label{fig5}
\end{figure}

Our goal is to create a programmable Heisenberg magnet $\hat{H}_{\text{H}}=\sum_{j>i}\mathcal{J}_{ij}\sum_{\alpha}\hat{\Lambda}_{\alpha}^{(i)}\hat{\Lambda}_{\alpha}^{(j)}$, where $\hat{\Lambda}_\alpha$ is the generalized Gell-Mann matrix (Appendix \ref{Appendix_Gell-mann}). The challenge of simulating SU($n$)-spin with cold atoms and ions is that the spin operators cannot be efficiently mapped to a rotation within an internal DOFs due to limited transition pathways, e.g., selection rules. In addition, there is a difficulty to implement spin-models with certain symmetries that cannot be imposed to the fundamental symmetries of the atomic interactions, e.g., SU($n$)-symmetric collisions in alkali-earth atoms limited by the nuclear spin DOF  \cite{Gorshkov2010}. Our method eliminates both bottlenecks, by locally encoding an ensemble of SU($2$) spins to the SU($n$)-subspace, and by building the interaction symmetry directly into the Hamiltonian in an emergent manner.

The general strategy is to impose an effective local gauge symmetry onto the spin system through the separation of time scale. We can then introduce a perturbative spin-exchange term that only virtually breaks the local symmetries. By construction, we aim to obtain a microscopic many-body dynamics within the gauge sector, which can be effectively interpreted as the macroscopic binary interactions between the SU($n$) spins. From the viewpoint of lattice gauge theories, the constrained quantum dynamics can be qualitatively understood as quantum fluctuations within the background gauge field of a frustrated vacuum of the logical spin system, which gives rise to a physical $4$-body plaquette interaction. 

As shown in Fig. \ref{fig5}, we partition the physical atomic lattice $i,j$ into logical spins $\overline{i},\overline{j}\in\mathcal{L}$, each containing $n$ physical atoms, that encode the local SU($n$) spin. This is achieved by local U($1$)-gauge constraints $\hat{G}_{\overline{i}}$ that blockade the total excitation number within the logical spin $\overline{i}$ to reside in the single-excitation subspace $\{|\alpha\rangle \equiv |s_{\alpha}\rangle\prod_{\beta \ne \alpha}|g_{\beta}\rangle\}$ with $\alpha\in \{1\cdots n\}$. Such a gauge generator $\hat{G}_{\overline{i}}=\sum_{i\in\overline{i}}\hat{\sigma}_z^{(i)}-{\mathcal{Q}}$ effectively imposes the Gauss law (``ice rules'') with electric charge ${\mathcal{Q}}={n-2}$, analogous to quantum spin ice models  \cite{Balents2010,Gingras2014} that mediate long-range ring-exchange interactions. The ground state (most excited state) sector of $\hat{H}_G=\lambda_G\sum_{\overline{i}}\hat{G}_{\overline{i}}^2$ for $\lambda_G>0 $ ($\lambda_G<0$) is spanned by $n$-dimensional states $\{|\alpha\rangle\}$ of the SU($n$)-representation. Without the loss of generality, we rewrite the Heisenberg model within this definition,
\begin{equation}
\hat{H}_{\text{H}}=\sum_{\overline{i}\ne \overline{j}}\mathcal{J}_{\overline{i},\overline{j}}\sum_{\alpha,\beta}\hat{\mathcal{T}}_{\alpha\beta}^{(\overline{i})}\hat{\mathcal{T}}_{\beta\alpha}^{(\overline{j})},\label{Heisenbergmagnetequation}
\end{equation}
where $\hat{\mathcal{T}}_{\alpha\beta}=|\alpha\rangle\langle\beta|$.

In order to introduce spin-spin interaction between the logical blocks, we treat the primitive Hamiltonian $\hat{H}_I=\sum_{\overline{i},\overline{j}}\hat{D}_{\overline{i},\overline{j}}+\hat{O}_{\overline{i},\overline{j}}$ as a perturbation to $\hat{H}_G$ with $\hat{D}_{\overline{i},\overline{j}}=D_{\overline{i},\overline{j}}\sum_{\alpha}\hat{\sigma}_{ss}^{(\overline{i}_\alpha)}\hat{\sigma}_{ss}^{(\overline{j}_\alpha)}$ and $\hat{O}_{\overline{i},\overline{j}}=O_{\overline{i},\overline{j}}\sum_{\alpha}\hat{\sigma}_{+}^{(\overline{i}_\alpha)}\hat{\sigma}_{-}^{(\overline{j}_\alpha)}$, where $\hat{\sigma}^{(i_\alpha)}$ denotes the spin operator acting on the $\alpha^{\text{th}}$ atom in the $i^{\text{th}}$ logical block. With the local gauge constraints, we obtain the effective Hamiltonian within the gauge-invariant sector $\mathcal{Q}$ as 
\begin{equation}
\hat{H}_\text{eff}=\sum_{\overline{i}\ne \overline{j}}D_{\overline{i},\overline{j}}\sum_{\alpha}\hat{\mathcal{T}}_{\alpha\alpha}^{(\overline{i})}\hat{\mathcal{T}}_{\alpha\alpha}^{(\overline{j})}+\mathcal{J}_{\overline{i},\overline{j}}\sum_{\alpha \ne \beta}\hat{\mathcal{T}}_{\alpha\beta}^{(\overline{i})}\hat{\mathcal{T}}_{\beta\alpha}^{(\overline{j})}, \label{PhysicalSU3equation}
\end{equation}
with the gauge-variant errors (spinon excitations) suppressed by the many-body gap $\lambda_G$ (spinon energy). In the physical space, the spin-exchange coefficients $D_{\overline{i},\overline{j}},\mathcal{J}_{\overline{i},\overline{j}}=-O^2_{\overline{i},\overline{j}}/2\lambda_G$ are the gauge-mediated ring-exchange interactions among the four spins selected by the primitive two-body model $\hat{H}_I$. With $D_{\overline{i},\overline{j}}=\mathcal{J}_{\overline{i},\overline{j}}$, the effective Hamiltonian is mapped to the universal SU($n$)-Heisenberg magnet $\hat{H}_{\text{H}}$. The gauge-projected Hamiltonian is derived in Appendix \ref{Appendix_GPHeisenberg}.

\begin{figure}[t!]
  \includegraphics[width=1\columnwidth]{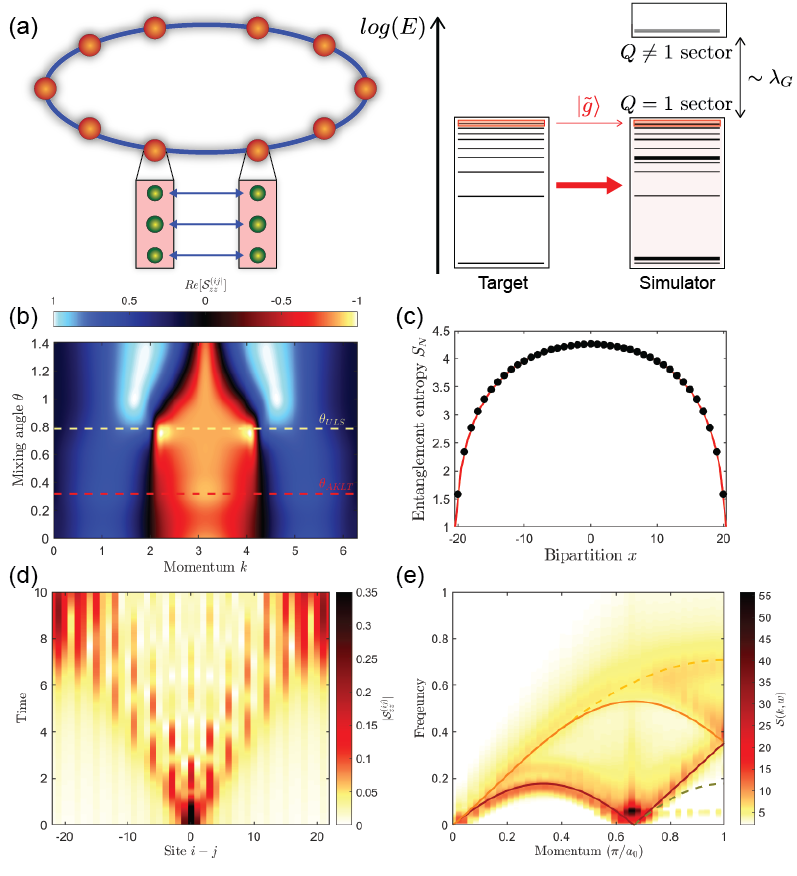}
  \caption{{Emergence of Wess-Zumino-Witten (WZW) conformal field theories (CFT).} (a) Local Hamiltonian encoding of SU($3$)$_{k=1}$ field theories on a ring onto SU($2$) waveguide QED simulator. The target WZW CFT is isometrically transformed to the local Hilbert space of the simulator with electric charge $Q=1$. (b) Phase diagram of the bilinear biquadratic spin-1 model with $N_{\text{eff}}=42$ logical blocks ($N=124$ atoms). Pinch points of static structure factor $\mathcal{S}^k_{zz}=\langle S_z^{k} S_z^{-k}\rangle$ at momentum $k=2\pi/3,4\pi/3$ signify the existence of divergent correlations at the Uimin-Lai-Sutherland (ULS) quantum critical point (QCP). The static structure factor is obtained from the correlation functions in Appendix \ref{Appendix_WZW} with uniform matrix product states (MPS) in the thermodynamic limit. (c) Critical scaling for entanglement entropy for vacuum state of (1+1)D SU($3$)$_k$ WZW field theory of level $k=1$. The vacuum entanglement entropy follows the Calabrese-Cardy formula for (1+1)D conformal field theories (CFT). The central charge $c=2.05\pm 0.03$ is extracted from the finite-size scaling. (d) Production of $c=2$ primary fields (quasiparticles) upon local quenching. Topological solitons carry fractional quantum statistics of Abelian anyonic phase $\phi=2\pi/3$. (e) Dynamical probes for quasiparticles of the WZW CFT. Ground states are obtained with a hybrid DMRG-TEBD algorithm for finite MPS in a complex-time coordinate (Appendix \ref{Appendix_WZW}). Dynamical structure factor is obtained by real-time evolving the ground state MPS with a TEBD algorithm.}
  \label{fig6}
\end{figure}

One feature of our synthetic approach is that the symmetries of the interaction can be directly built into the underlying Hamiltonian, without resorting to the fundamental symmetries of the atomic collisions. For instance, with a minor modification, we can easily create SU($n$)-symmetric Hamiltonians for arbitrary $n$, e.g., unlimited by the nuclear-spin DOF, for the study of transition metal oxides  \cite{Tokura2000} and heavy fermion systems. Furthermore, because we can design $\mathcal{J}_{\overline{i},\overline{j}}$ arbitrarily through the Raman fields, our system can be tailored to study novel frustrated magnetic ordering in long-range SU($n$)-spin models with the Haldane gap \cite{Read1989,Marston1989,Greiter2007,Affleck1989}. As discussed in the next section, our waveguide QED simulator can be applied to the realization of quantum field theories \cite{Wess1971, Witten1984}.

In Appendix \ref{Appendix_SYModel}, we discuss an efficient method to construct the real-time evolution of the Sachdev-Ye (SY) model \cite{Sachdev1993} with dynamical Raman fields, an all-to-all $\lim_{n\rightarrow \infty}$SU($n$)-Heisenberg model $\hat{H}_{\text{SY}}$ (Eq. \ref{Heisenbergmagnetequation} with Gaussian-random $\mathcal{J}_{\overline{i},\overline{j}}$). The SY model describes a non-Fermi liquid state of matter, known as the ``strange metal,'' characterized by the absence of long-ranged quasiparticle excitations analogous to high-$T_c$ cuprate superconductors. In connection to quantum chaos \cite{Sachdev2015}, a quenched system under $\hat{H}_{\text{SY}}$ rapidly loses the phase coherences and reaches a quantum many-body chaos within time scales that remarkably saturate the quantum bound of the Lyapunov time $\tau_L=\frac{\hbar}{2\pi k_B T}$. With gauge-mediated many-body string Hamiltonian between a set of SU$(n)$-spins and an ancilla qubit, we can even directly assemble and measure arbitrarily complex OTOCs  \cite{Kitaev2014,Shenker2014, Maldecena2016} $\langle  \hat{W}^\dagger(\tau)\hat{V}^\dagger(0)\hat{W}(\tau)\hat{V}(0) \rangle\sim e^{\tau/\tau_L} $ for SU($n$)-variables $\hat{W},\hat{V}$ in our platform for the detection of the quantum chaos and the scrambling of entanglement in many-body quantum systems. The SY model also serves as a model of holography that duals quantum gravity in AdS$_2$/CFT  \cite{Sachdev1993,Sachdev2015}.

\section{Strongly-coupled WZW field theory}\label{WZW_section}

Quantum field theories (QFT), defined on continuous spacetimes with each site supporting infinite-dimensional Hilbert spaces, become increasingly intractable to simulate in the regime of strong coupling even on quantum devices. Near the strong coupling, the physics of the UV fixed point is often described by conformal field theories (CFT) with a scale-invariant and universal description. Moreover, extracting the conformal data of the emergent CFT is a notoriously difficult task for real quantum hardwares. In an examplary fashion, we demonstrate the emergence of (1+1)D SU($n$)$_k$ Wess-Zumino-Witten (WZW) CFT \cite{Wess1971,Witten1983,Witten1984} in the waveguide-coupled SU($n$) Hamiltonian (Fig. \ref{fig6}(a)), which describes the boundary physics of a bulk ($2+1$)D Chern-Simons topological gravity in the scaling limit \cite{Witten1989}. In condensed matter systems, WZW theory serves as the parent that hosts a family of symmetry-protected gapless edge states in fractional quantum Hall systems. The primary fields $\Psi$ of the CFT are produced and monitored by way of real optical fields of the guided modes. The long-wavelength conformal data, including the central charge $c$, the quantum dimensions $\mathcal{D}$, and operator product expansion of $\Psi$, is reconstructed from the correlation between physical observables of the microscopic simulator, as reflected by the fluctuation of the optical fields in the guided mode. 

As discussed in Fig. \ref{fig6}(a), we consider a critical SU($3$)-Heisenberg Hamiltonian for nearest-neighbor interacting $N_\text{eff}$ logical SU($3$) spins living on a ring with $\mathcal{J}_{\overline{i},\overline{i}+1}=\mathcal{J}_c$ for Eq. \ref{Heisenbergmagnetequation} (See the phase diagram of Fig. \ref{fig6}(b) with quantum critical point $\theta_{\text{ULS}}=\pi/4$). The target system is mapped to the waveguide QED simulator (i) by creating nearest-neighbore bonds between physical atoms with $\hat{H}_I$ (blue arrows of Fig. \ref{fig6}(a)) and (ii) by gauging the simulator to $\mathcal{Q}$ (red shaded area of  Fig. \ref{fig6}(a)). The gauged spectra of the simulator (with $\lambda_G$) is thereby that of the target with an error $(D/\lambda_G)^2\ll 1$. To access the ground state $|\tilde{g}\rangle$ of the target model (most excited state of the gauged simulator), we perform a hybrid matrix-product state (MPS) algorithm for the waveguide quantum simulator moving along a complex time, combining both density-matrix renormalization group (DMRG) and time-evolving block decimation (TEBD) methods. By evolving a random MPS under the action $\lim_{t\rightarrow\infty}\exp\left[-i (\hat{H}_I+i\hat{H}_G) t\right]$, we obtain the most excited state within the low-energy sector $\mathcal{Q}=1$ of the simulator, which is isometric to the DMRG ground state of the logical antiferromagnetic SU($3$) model (Appendix \ref{Appendix_WZW}). 

To see how the SU($3$)$_1$ WZW CFT for level $k=1$ natively emerges from the Hamiltonian constraints of the simulator, let us consider the parton picture of the target Hamiltonian (See Eq. \ref{Heisenbergmagnetequation}). We map the logical operators with $3$-color fermions (quarks) $\hat{\mathcal{T}}_{\alpha\beta}^{(\overline{i})}=\hat{\psi}^{(\overline{i})\dagger}_{\alpha}\hat{\psi}^{(\overline{i})}_{\beta}$ under the constraint $\hat{\psi}^{(\overline{i})\dagger}_{\alpha}\hat{\psi}^{(\overline{i})}_{\alpha}=1$ for colors $\alpha,\beta=\{r,g,b\}$. The parton Hamiltonian 
\begin{equation}
\hat{\mathcal{H}}_{\text{parton}}=\mathcal{J}\sum_{\overline{i}} \hat{\psi}^{(\overline{i})\dagger}_{\alpha}\hat{\psi}^{(\overline{i})}_{\beta}\hat{\psi}^{(\overline{i}+1)\dagger}_{\beta}\hat{\psi}^{(\overline{i}+1)}_{\alpha}\label{SUNsectionpartonEq}
\end{equation}
is equivalent to a SU($3$) Hubbard model $\mathcal{H}_{\text{Hubbard}}=-\sum_{\overline{i}}t[\hat{\psi}^{(\overline{i})\dagger}_{\alpha}\hat{\psi}^{(\overline{i}+1)}_{\alpha}$ $+h.c]+U[\hat{\psi}^{(\overline{i})\dagger}_{\alpha}\hat{\psi}^{(\overline{i})}_{\alpha}-1]^2$ for fermions in the interaction limit $U/t\gg 1$. In the infrared, low-energy excitations are only populated at the Fermi points $k_F=\pi/3$, thereby coarse-graining the fermionic fields $\hat{\psi}^{(\overline{i})\dagger}_{\alpha}=e^{i k_F x_{\overline{i}}}\hat{\psi}_{L,\alpha}(x_{\overline{i}})+e^{-i k_F x_{\overline{i}}}\hat{\psi}_{R,\alpha}(x_{\overline{i}})$ to the continuum. As the Hubbard model for $U/t\ll 1$ gives rise to $3$-color free Dirac fermions (charge boson and SU($3$)$_1$ WZW gauge fields $g$), the Hubbard interaction asymptotically decouples the charge with a gap. Thus, the Hubbard interaction leaves the WZW fixed point in the low-energy sector with an action $  S=\frac{1}{16\pi}\int_{\mathcal{G}^2} d^2\xi \text{Tr}[\partial_{\alpha} g^{-1}\partial^{\alpha} g]+\Gamma(g)$
and topological term $\Gamma(g)=\frac{1}{24\pi}\int_{\mathcal{G}^3} d^3 \xi \epsilon^{\alpha\beta\gamma}\text{Tr}[ (g^{-1}\partial_{\alpha} g)(g^{-1}\partial_{\beta} g)(g^{-1}\partial_{\gamma} g)]$. This is reminiscent to chiral Luttinger liquids (LL) on fractional quantum Hall edges \cite{Sule2015}. Unlike the Haldane phase of the spin-$1$ counterpart, the emergent field theory of the SU($3$) model is described by universal properties, where the (chiral) fermionic fields $\hat{\psi}_{L,\sigma},\hat{\psi}_{R,\sigma}$ become the Virasoro primary fields $g_{\alpha\beta}(z,\overline{z})=\hat{\psi}_{L,\sigma}^{\dagger}(z)e^{i\hat{\phi}(z,\overline{z})}\hat{\psi}_{R,\sigma}(\overline{z})$ of the WZW CFT with colors $\sigma=\{r,g,b\}$ and space-time $z=-i(x-t),\overline{z}=i(x+t)$. These fields are generated by the spin currents $J^{a}_L (x)=\frac{1}{2}\hat{\psi}_{L,\sigma}^{\dagger}(x)\tau^{a}_{\sigma,\sigma^{\prime}}\hat{\psi}_{L,\sigma^{\prime}}(x)$ and $J^{a}_R (x)=\frac{1}{2}\hat{\psi}_{R,\sigma}^{\dagger}(x)\tau^{a}_{\sigma,\sigma^{\prime}}\hat{\psi}_{R,\sigma^{\prime}}(x)$ following the SU($3$)$_1$ Kac-Moody algebra, where $\tau^{a}_{\sigma,\sigma^{\prime}}=(\Lambda_a)_{\sigma,\sigma^{\prime}}/2$ are the elements of the generalized Gell-Mann matrices in Appendix \ref{Appendix_Gell-mann}. Importantly, from the operator product expansion, the conformal data of SU($3$)$_1$ WZW CFT can be obtained for the central charge $c=2$, scaling dimensions $\mathcal{D}=\frac{2}{3}$ and critical exponents $\nu=2$ for the WZW field $g_{\alpha\beta}$.

In order to physically extract the conformal data from the simulator, we need to measure the static and dynamic response functions. To this end, we dissipate an observable $\hat{\mathcal{O}}^{(j)}$ of the physical atom at site $j$ to the waveguide $\sum_{\overline{j}}g_{\overline{j}}\sum_{j\in\overline{j}} \hat{\mathcal{O}}^{(j)}\hat{a}_{k_0} e^{i k \overline{j}}$ with a well-defined momentum $k$. The first-order correlation $\langle\normord{\hat{a}^{\dagger}(\tau)\hat{a}(0)}\rangle_k$ of the optical field leaving the guided mode regresses towards the dynamical response function $\mathcal{S}_{\mathcal{O}}^k (\tau)=\langle 0_{\text{CFT}}|\hat{\mathcal{O}}_{-k} (\tau) \hat{\mathcal{O}}_{k}(0)|0_{\text{CFT}}\rangle$ of the logical spin system, where $|0_{\text{CFT}}\rangle$ is the vacuum state of the WZW CFT and $\hat{\mathcal{O}}_{k}=\sum_{\overline{j}}\frac{g_{\overline{j}}}{\kappa_0}\sum_{j\in\overline{j}} \hat{\mathcal{O}}^{(j)}e^{i k \overline{j}}$. This method allows us to construct a broad class of static and dynamical structure factors of the many-body system, giving access to the low-energy excitations as well as the universal properties $\nu$ and $\mathcal{D}$ of the CFT. In Appendix \ref{Appendix_WZW}, we analyze our result for the spin correlators $\langle \hat{S}_z^{(\overline{i})} \hat{S}_z^{(\overline{j})} \rangle\sim |\overline{i}-\overline{j}|^{-2\mathcal{D}}$ and extrapolate the scaling dimensions $\mathcal{D}=0.67\pm0.02$ with the DMRG ground state up to $N_{\text{eff}}=200$ logical blocks. We also characterize the correlation length $\xi \sim |\theta-\theta_c|^{-\nu}$ with the critical exponents $\nu=2.10\pm 0.05$ for the bilinear biquadratic (BBQ) spin-$1$ Hamiltonian with Uimin-Lai-Sutherland (ULS) quantum critical point (QCP) $\theta_c=\pi/4$ with an enlarged SU($3$)-symmetry, corresponding to our SU($3$) Heisenberg model. 

In Fig. \ref{fig6}(b), we present the phase diagram of the BBQ model detected with the static spin structure factor $\mathcal{S}_{zz}^{k}=\langle \hat{S}_z^{-k}\hat{S}_z^{k}\rangle$. Near the ULS QCP, power-law singularities appear in the form of pinch points at the momentum $k=2\pi/3$ and $4\pi/3$, indicative of absence of long-range order (disordered state) for the SU($3$) spin model and the gapless soliton excitations on top of the CFT vacuum (algebraic spin liquid state). These topological solitons appear to carry anyonic statistics with Abelian phase $\phi=\pi$. Upon locally quenching the many-body system with $\hat{S}_z^{(i)}$, these solitons can be produced in pairs moving at the Fermi velocity $v_F=\pi/3$ (Fig. \ref{fig6}(d)). To assess the spectral properties of WZW fields, we probe the dynamical structure factor $\mathcal{S}_{zz}(w,k)=\int \exp(i w\tau) \mathcal{S}_{zz}^{k}(\tau)$ in Fig. \ref{fig6}(e). Two soliton modes are visible in the contour map $\mathcal{S}_{zz}(w,k)$ (See the two solid guiding lines), reflected by their length scales $3/2\pi$ and $3/4\pi$. In addition, the solitonic continuum appears smoothly as the quisiparticle populations between the two solid lines due the coherence between the soliton pairs, and higher order $4$-local soliton excitations begin to appear between the dashed black line and the solid red line for $k>2\pi/3$. 

We characterize the central charge $c$ of the CFT by scaling the entanglement entropy $S=-\text{Tr}[\rho_A \ln \rho_A]$ between the subsystems $A$ and $B$ of the logical system with $\rho_A=\text{Tr}_B |0_{\text{CFT}}\rangle\langle 0_{\text{CFT}}|$. In the framework of entanglement Hamiltonian $\tilde{H}_A=\sum_l \tilde{\epsilon}_l|\tilde{\epsilon}_l\rangle\langle \tilde{\epsilon}_l |$, we consider the problem of extracting thermodynamic property of the state $\rho_A=\exp (-\tilde{H}_A)=\sum_{l}e^{-\tilde{\epsilon}_l}|\tilde{\epsilon}_l\rangle\langle \tilde{\epsilon}_l |$, where $\{\tilde{\epsilon}_l\}$ is the entanglement spectrum for the CFT vacuum state $|0_{\text{CFT}}\rangle$ \cite{Dalmonte2018}. The entanglement entropy $S=\sum_l \tilde{\epsilon}_l \ln (\tilde{\epsilon}_l)$ is then obtained from the entanglement Hamiltonian $\tilde{H}_A$ at an effective temperature $T=1$, whose eigenspectrum $\{\epsilon_l\}$ is determined by many-body spectroscopy \cite{Li2008,Senko2014}. Importantly, due to the Bisognano-Wichmann theorem, the entanglement Hamiltonian $\tilde{H}_A$ can be cast in terms of the original model $\hat{H}_H$ (See Eq. \ref{Heisenbergmagnetequation}) with inhomogeneous coupling $\mathcal{J}_{\overline{i},\overline{i}+1}=\mathcal{J}_c \Gamma(\overline{i})$ and prefactor $\Gamma(x)=\frac{N_{\text{eff}}}{\pi}\sin\left (\frac{\pi x}{N_{\text{eff}}}\right)$ defined over a subsystem $\overline{i}\in A$ \cite{Dalmonte2018}, which can be simulated by the SU($N$) toolbox of Eq. \ref{PhysicalSU3equation}.

In Fig. \ref{fig6}(c), we present our result of the entanglement entropy for SU($3$) Heisenberg model. At quantum critical points in $(1+1)$D, the vacuum-state entanglement entropy $S(x,N_{\text{eff}})$ logarithmically scales with the system size $N_{\text{eff}}$, following the Calabrese-Cardy formula $S(x,N_{\text{eff}})=\frac{c}{3}\log\left [ \left(\frac{N_{\text{eff}}}{\pi}\sin\left( \frac{\pi x}{N_{\text{eff}}}\right)\right)\right]$ for the bipartite cut $x$ up to a non-universal offset \cite{Calabrese2009}. By fitting to the Calabrese-Cardy formula, we thereby obtain the central charge $c=2.05\pm0.03$, consistent with the CFT prediction $c=2$. In the parton theory, the $c=2$ WZW CFT is manifested by the two-component non-Abelian bosons of the Luttinger liquid. The tensor product of compactified bosons, each carrying $c=1$ in the dual space, effectively give rise to the $c=2$ field theory for the SU($3$)-symmetric spin model.

As a final remark, while this section focused on the minimal instance $(1+1)$D SU($3$)$_1$ WZW CFT, our waveguide QED simulator and meausrement protocols are directly applicable to a wider class of WZW CFT. Namely, the symmetry group SU($n$)$_k$ and level $k$ can be engineered with the local encoding $n=N/N_{\text{eff}}$ and the sector $\mathcal{Q}$, and long-range interactions can be introduced for arbitrary spatial dimensions. Unlike the Abelian-like spin liquids described by SU($3$)$_1$ WZW CFTs, SU($3$)$_k$ WZW CFTs are genuinely interacting CFTs, and host a far richer family of non-Abelian anyons.

\section{Discussion}\label{discussion_section}

Realization of universal quantum matter with waveguide QED simulator presents technological challenges which can be addressed by state-of-art nanophotonic experiments \cite{Tiecke2014,Goban2014,Goban2015,Hood2016}. Defect-free atomic arrays can be generated in free-space with acousto-optical deflectors \cite{Barredo2016,Endres2016} and spatial light modulators \cite{Kim2016}. With evanescent cooling and advanced side-illumination loading techniques for PCW structures \cite{Meng2017,Hung2013}, it is conceivable to prepare defect-free atom array on flat-band PCWs, such as the SPCW. In Appendix  \ref{Appendix_SPCWsection}, we provide an example of a SPCW tailored to achieve the desired photonic bands for renormalizing individual Cs atoms to universal quantum matter. Programmable control of the exchange coefficients requires the capability to tune $\sim N^2$ phase-amplitudes of the Raman sideband matrices in tandem. Such a capability has been adapted for $100$-spin coherent Ising machine \cite{McMahon2016,Haribara2017}, and ultrafast multimode modulation techniques have been developed in the telecommunication industry.

In the waveguide QED simulator, the correlated Lamb shift in the PCW generates mechanical interaction between the external motional states of the trapped atoms. In turn, the Bogoliubov phonons are exploited as a quantum bus for mediating the universal Hamiltonian. Compared to other networked quantum architectures, the PCW allows a versatile control over both the dissipative loss and coherent dispersion (single-particle band structure) of such a bus. The figure of merit $\mathcal{F}$ provides a natural scaling parameter for the coherence-to-dissipation ratio of the simulator. In the reactive regime, $\mathcal{F}\sim m_e\exp({L_d/L_c}) \gg 1$ exponentially improves with longer device length $L_d$ for a given photon mass $m_e$. As an example, in Appendix \ref{Appendix_SPCWsection}, we numerically simulate the Green's function $G_\text{1D}(\hat{{x}}_i,\hat{{x}}_j)$  and find $\mathcal{F}\sim 10^4$ for the Silicon Nitride SPCW structure.

Waveguide QED offers a unique playground for neutral atoms, in which light, motion and spin are all intertwined by the electromagnetic vacuum of the dielectric. By engineering the coupling between the phononic superfluid and the atomic spins, we have provided an analogue framework for simulating universal quantum matter with cold atoms. Such a simulator can be applied for universal quantum computation with continuous-time quantum cellular automata and Hamiltonian quantum computation \cite{Biamonte2008,Nagaj2008,Vollbrecht2008}. Our waveguide QED simulator utilizes largely non-interacting phonons with $L_c\gg x_0$. In the limit $L_c\simeq x_0$, the kinetic term of the extended Bose-Hubbard model $\hat{H}_{\text{M}}$ is constrained by the density-density interaction. Under such local gauge symmetries, complex lattice gauge theories beyond truncated quantum link models can emerge from the coherent coupling between the spin matter and fluctuating gauge phonons, renormalizing ordinary non-interacting matter to quantum field theories with the waveguide dielectric.

\begin{acknowledgments}
S. Maurya and A. Boddeti have contributed to the numerical simulation of the PCWs. S. G. Weiss at the IQC IST team has provided support to the computational cluster. This work is supported by the NSERC, the Canada Foundation for Innovation, the Ontario Ministry of Research and Innovation, NVIDIA, the KIST Institutional Program, the Compute Canada, and the Industry Canada. YD was supported by the Major Scientific Research Project of Zhejiang Lab (2019MB0AD01). YSL was supported by the Mike and Ophelia Lazaridis Fellowship Program.
\end{acknowledgments}

\appendix

\section{PERFECT TRANSFER IN A SPIN CHAIN}\label{Appendix_QST}
  
  To benchmark and verify the various approximations made for Eq. \ref{twobodyHam}, we simulate an 1D quantum wire that enables perfect quantum-state transfer (QST) between remote spin registers \cite{Bose2003,Christandl2004,Paternostro2005,Yung2005,DiFranco2008, Yao2011, Yao2013}. In particular, we compare the effective dynamics of $  \hat{H}_{\text{QST}}$ to that of the parent Hamiltonian $\hat{H}=\hat{H}_M+\hat{H}_A+\sum_{i,j}\sum_{\alpha,l}\frac{\Omega^{(j)}_{\alpha,l}}{2}\hat{\sigma}_{\alpha}^{(i)}\sin(k^{(j)}_{{\alpha},l}\hat{x}_{i})e^{-i\nu^{(j)}_{{\alpha},l}t}+h.c$ in Section \ref{Lamb_section}. We prepare an 1D spin medium with the translationally-variant XX Hamiltonian
  \begin{equation}
  \hat{H}_{\text{QST}}=\sum_{i=1}^{N-1}\frac{J^{(i,i+1)}}{2}(\hat{\sigma}_{x}^{(i)}\hat{\sigma}_{x}^{(i+1)}+\hat{\sigma}_{y}^{(i)}\hat{\sigma}_{y}^{(i+1)}),\label{qstequation}
  \end{equation}
  where $J^{(i,i+1)}=\alpha\sqrt{i(N-i)}$ and $\alpha$ is a global interaction constant. We solve the system parameters $\{\Omega^{(i)}_{x,k},\Omega^{(i)}_{y,k}\}$ from the set of nonlinear equations for $J^{(i,i+1)}$ under the constraint of minimum total intensity $\sum_{i,l} (|\Omega^{(i)}_{x,l}|^2+|\Omega^{(i)}_{y,l}|^2)$.
  
\begin{figure}[t!]
    \includegraphics[width=1\columnwidth]{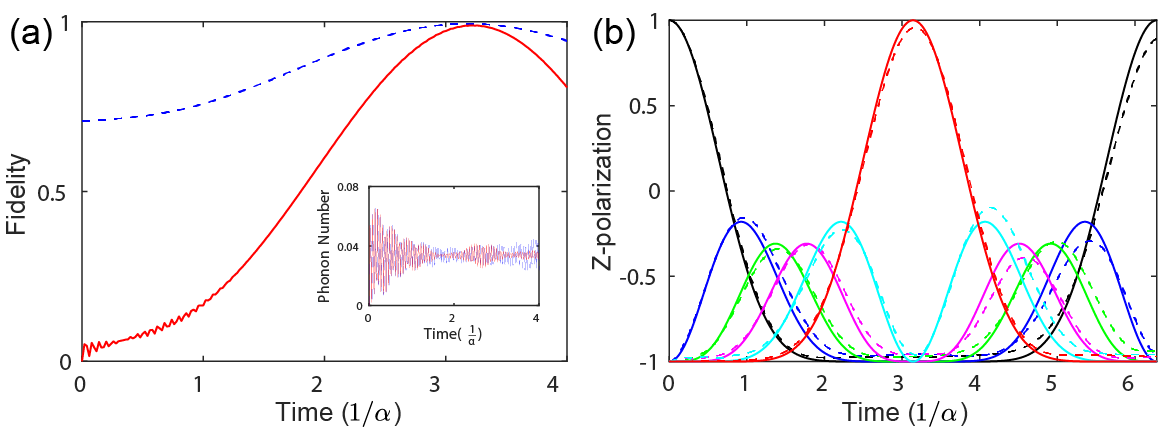}
    \caption{{Quantum-state transfer over a spin chain.} (a)  Fidelity between the real-time state on the last spin and the initial state on the first spin for two input states $|\psi_{\text{in}}^{(1)}\rangle=(|g\rangle-|s\rangle)/\sqrt{2}$ (red line) and  $|\psi_{\text{in}}^{(2)}\rangle=|s\rangle$ (blue dashed line). Inset is the mean number of phonons with a maximum value about $0.06$, which shows that phonon is rarely populated in the whole process and validates the adiabatic elimination of phonons. The dynamics is numerically simulated for the full Hamiltonian, which includes the interactions of the atomic internal states, phonons, and electromagnetic vacuum. Close-to-unit fidelity $F=0.994$ is achieved over time scale $t_f\simeq \pi/\alpha$. (b) Real-time dynamics of spin polarization $\langle\hat{\sigma}_z\rangle$ for all sites on the chain. The dashed (solid) line is obtained from the full (effective) Hamiltonian (in Eq. \ref{qstequation})}\label{QSTfig}
    \end{figure}

  As discussed in Ref.   \cite{Bose2003}, $\hat{H}_{\text{QST}}$ achieves the perfect state transfer of arbitrary input states $|\psi_{\text{in}}\rangle$ between the edge sites $i=1,N$ over arbitrarily long $N$ with unit fidelity by virtue of the mirror symmetry in the spin-exchange coefficients $J^{(i,i+1)}$. Unlike sequential direct state transfer, no external manipulation or feedback on the spin chain is required, and the complete transfer is achieved within transfer time $t_f=\pi/\alpha$ without state-preparation of the global spin chain. In Fig. \ref{QSTfig}(a), we simulate the full Hamiltonian dynamics of quantum-state transfer for two input states $|\psi_{\text{in}}^{(1)}\rangle=(|g\rangle-|s\rangle)/\sqrt{2}$ (red line) and  $|\psi_{\text{in}}^{(2)}\rangle=|s\rangle$ (blue dashed line) through an 1D atomic chain with $N=6$ atoms without eliminating the phonon fields. We  keep the coupling terms between those mismatched sidebands and Bogoliubov phonon modes. By sampling various input states coupled to an initially polarized spin medium, the minimal QST fidelity for pure states is numerically determined $F=\text{Tr}[|\psi_{\text{in}}^{(1)}\rangle\langle \psi_{\text{in}}^{(1)} |\rho_s]=0.994$ at $t_f \simeq \pi/\alpha$, yielding only $0.5\% $ error in the final state, testifying the accuracy of the effective Hamiltonian $\hat{H}_{\text{QMA}}$ in Eq. \ref{twobodyHam}.

  As shown in the inset of Fig. \ref{QSTfig}(a), the phonons across the entire spin chain are hardly populated throughout the state transfer, justifying the adiabatic elimination procedure. In Fig. \ref{QSTfig}(b), we also compare the full atom-phonon dynamics (solid lines) of the individual spin-polarizations $\langle \hat{\sigma}_z^{(i)}\rangle$ for an initially polarized spin medium $|g\cdots g\rangle$ with that of the reduced two-body Hamiltonian $\hat{H}_{\text{QST}}$ in Eq. \ref{qstequation} (dashed lines). When $|\psi_{\text{in}}^{(2)}\rangle$ is injected to the first spin (black line), the spin-excitation delocalizes across the entire spin chain and coherently builds up the its amplitude at the final spin with $\langle \hat{\sigma}_z^{(6)}\rangle\simeq 1$ at $t\simeq \pi/\alpha$ (red line). The minute difference between the solid and dashed lines affirms the various approximations for $\hat{H}_{\text{QMA}}$.  
  In Appendix \ref{Appendix_SPCWsection}, we simulate the full open-system dynamics of QST for Eq. \ref{qstequation}, by starting from the Green's tensor $\boldsymbol{G}(\boldsymbol{x},\boldsymbol{x}^\prime,w)$ of the candidate PCW structure in Fig. \ref{fig1}, and incorporate all known dissipative mechanisms intrinsic to our protocol. Such an effective dynamics is shown to be immune from the structural disorders of the PCW at the tolerance levels of state-of-art nanofabrication  \cite{Yu2014}.

\begin{figure*}[t!]
  \includegraphics[width=1.5\columnwidth]{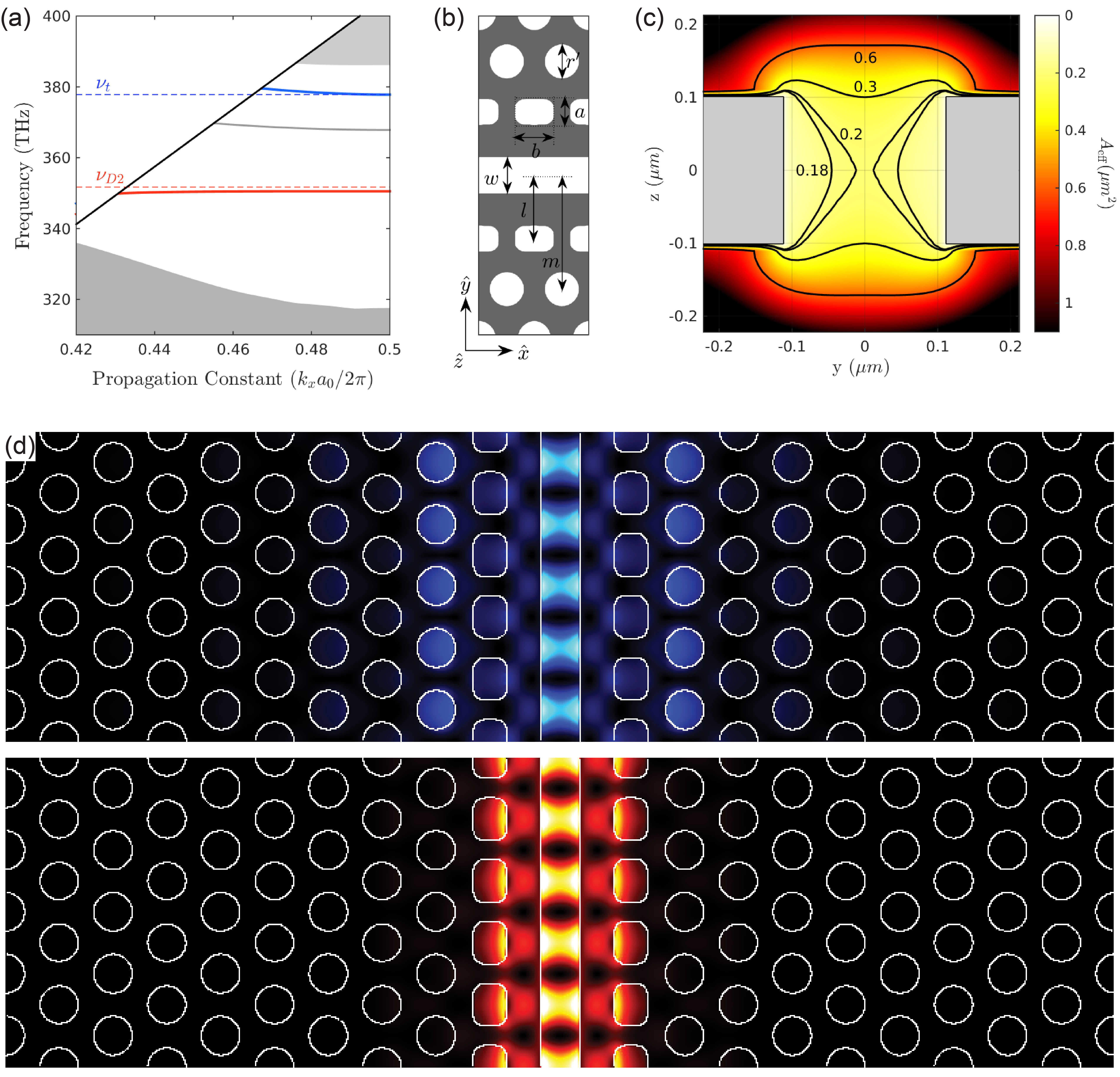}
  \caption{{Slotted squircle photonic crystal waveguide.} (a) SPCW band diagram. The guided modes are depicted as solid lines for both the excitation $\nu_{D2}$ (red) and trapping modes $\nu_{t}$ (blue). Through our optimization iterations, the guided modes (GM)  $\nu_{D2},\nu_{t}$ are flattened around the Cesium $D_2$-transition and magic-wavelength trapping frequencies. GM $\nu_{t}$ is defined to operate at the blue-detuned magic wavelength condition for the $D_2$-transition at $\lambda_t=793.5$ nm. The grey shaded region indicates the presence of slab modes. (b) SPCW geometry. The parameters that define the SPCW structure is provided in Table \ref{SPCWtable}. (c) Effective mode area $A_{\text{eff}}$. We depict the x-cut contour map of $A_{\text{eff}}$ for GM $\nu_{D2}$. At the trapping region, we anticipate sub-wavelength localization $A_{\text{eff}}/\lambda_{D2}^2\simeq 0.18$ and effective coupling rate $g_c\simeq 11.5$ GHz. The resulting photonic Lamb shift and localization length are $\Delta_{\text{1D}}\simeq 620$ MHz and $L_c\simeq 0.77$ $\mu$m at $\Delta_e=0.4$ THz. (d) Contour intensity map of the guided modes $\nu_{D2},\nu_{t}$.}\label{bandfig}
  \end{figure*}

  \section{SQUIRCLE PHOTONIC CRYSTAL WAVEGUIDE}\label{Appendix_SPCWsection}
  
  \begin{figure*}[th!]
  \includegraphics[width=1.75\columnwidth]{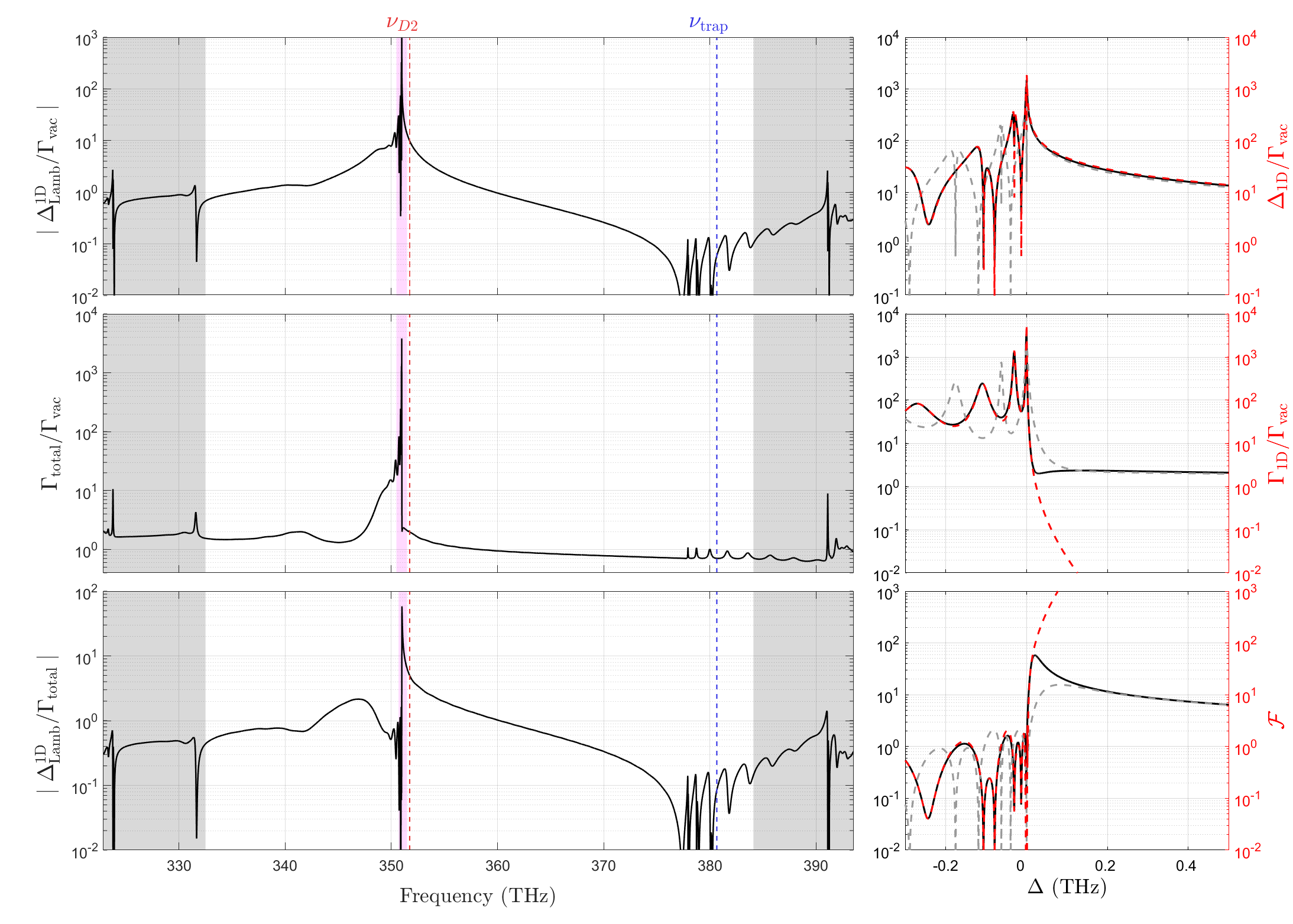}
  \caption{{Collective atomic decay and photonic Lamb shift of a finite SPCW.} (a) Photonic Lamb shift $\Delta^{\text{1D}}_{\text{Lamb}}$ for electronically excited states. The energy shift $\Delta_{\text{1D}}$ of the excited state $|6P_{3/2},F=4\rangle$ of Cs is computed by the numerically evaluating the local scattering Green's function $\boldsymbol{G}_s(\boldsymbol{x},\boldsymbol{x}^\prime,\omega)$. We only consider the level shift caused by the SPCW structure, but not the absolute renormalization by the electromagnetic vacuum. As a benchmark, we normalized the Lamb shift by the free-space decay rate $\Gamma_{\text{vac}}$. We also display the photonic Lamb shift $\Delta_{\text{1D}}$ under the single-band approximation as red dashed line. The close agreement between the two models testify the accuracy of the extrapolated $\Gamma_{1D}$. (b) The enhancement and inhibition of spontaneous emission in dispersive and reactive regimes. The total decay rate $\Gamma_{\text{total}}$ is strongly enhanced at the band edge, and is exponentially inhibited in the band gap with $\Gamma_{\text{total}}\simeq \Gamma_{\text{1D}}\exp (-L_d/L_c)$, where $\Gamma_{\text{1D}}$ is the enhanced decay rate at the resonance closest to the band edge, $L_d=80a_0$ is the device length for lattice constant $a_0$, and $L_c$ is the localization length. Deep into the band gap $\Delta_e\gg 0$, the reduction of $\Gamma_{\text{total}}$ is limited by the weakly inhibited homogeneous decay rate $\Gamma^\prime\simeq 0.7 \Gamma_{\text{vac}}$ that predominantly emits photons out of plane of the slab. (c) Lamb shift to decay rate ratio $\Delta_{\text{1D}}/\Gamma_{\text{total}}$ across a wide detuning range up to $\Delta_e\simeq 10$ THz. Inset. Figure of merit $\mathcal{F}\gg 1$ (red dashed line). The grey shaded region indicates the presence of slab modes.}\label{J1DGammafig}
  \end{figure*}
  
 The full realization of our waveguide QED toolboxes requires the capability to maintain favourable figure of merit $\mathcal{F}=\Delta_{\text{Lamb}}/\Gamma_{\text{tot}}$ with short-ranged mechanical interactions between the trapped atoms, where the localization length $L_c=\sqrt{1/2m_e \Delta_e}$ is comparable to the lattice constant $a_0$. Here, $\Delta_e\simeq 2\Delta_b$ denotes the detuning of atomic transition to the effective cavity mode \cite{Douglas2015}, and $\Delta_b$ is the detuning of the atomic transition frequency to the band edge.  While it is not necessary to have nearest-neighbour interactions with sparse loading, the atomic collective motion can experience band-flattening effect due to the long-range phonon tunnelling, which reduces the local addressability of the spin-motion couplings. For laser cooling and trapping nearby the nanoscopic structures, the PCW requires a wide angular field of view for the optical access, and restrict the dimensions of PCW slabs to 1D and 2D. Because of the lack of full 3D PBGs, the total decay rate  $\Gamma_{\text{tot}}=\Gamma_{\text{1D}}+\Gamma^{\prime}$ consists of both the waveguide decay $\Gamma_{\text{1D}}$ and the homogeneous decay $\Gamma^{\prime}$. While  $\Gamma_{\text{1D}}$ is significantly suppressed for large $\Delta_e$, majority of slow-light PCWs do not have the adequate band structure with large $m_e$ to induce strong coherent motional coupling with $\mathcal{F}\gg1$ at small $L_c$.
  
  \subsection{System parameters}
  In this section, we discuss a variation of a slotted PCW that utilizes PBG of the 2D slab as the guiding mechanism  \cite{Krauss2007,Arcari2014,Lodahl2015}. As shown in Fig. \ref{bandfig}, the dispersion is tailored by a line defect introduced to a triangular TE-PBG slab, where a significant portion of the energy of the GM is localized within the air slot. We introduce anomalous squircles in the vicinity of the air slots to alter their band curvatures. The rationale of our dispersion engineering is that the combination of the lattice constant $a_0$, the hole radius $r$, and the air slot width $w_s$ can tune the locations of the band edge frequencies with respect to the band gap of the slab, while the additional squircle geometries defined by the asymmetry $a, b$ cause differential energy shifts between the z-even bands of opposite x-symmetry. By placing the bands deep into the PBG of the surrounding slab, we suppress the $k$-space interval $[k_c, k_l]$ where the in-plane field profile of the GM is localized by index-guiding near the light cone. The proximal squircle geometry then flattens the GM across the band-gap guided $k$-space fraction $[k_l, k_0]$. In addition, the out-of-plane emission $\Gamma_{h}$ is affected by the distance of the squircles to the slot.

  \begin{figure*}[t!]
    \includegraphics[width=1.5\columnwidth]{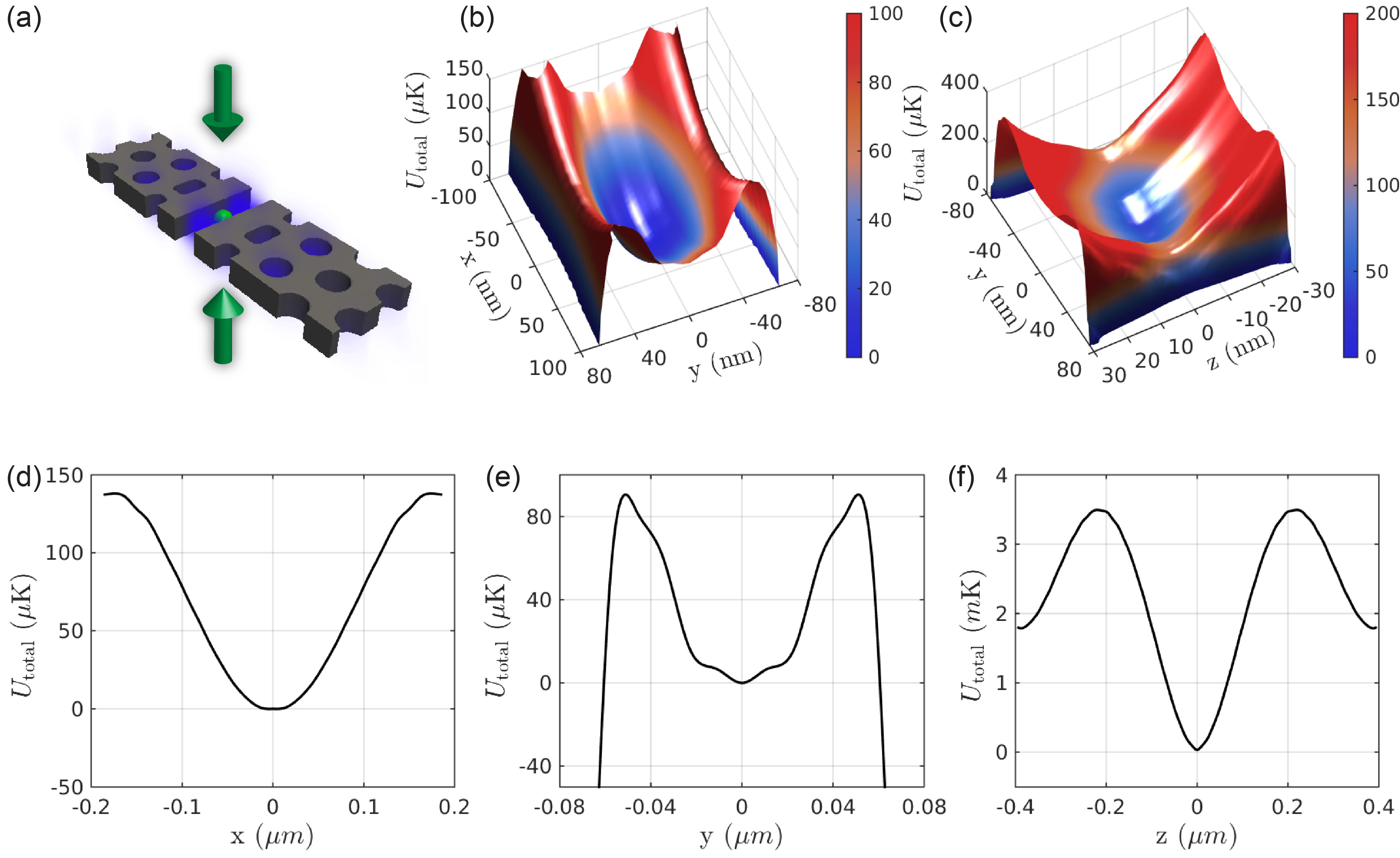}
    \caption{{Adiabatic ground-state potentials for Cesium atom assisted by side-illumination beams.} Cesium trapping potentials of $|6S_{1/2}\rangle$ for (b) $x-y$ plane and (c)  $y-z$ plane with (d) the $x$-, (e) $y$-, and (f) $z$-slices. We assume that the refractive index $n$ is frequency-independent. The coordination system $(x,y,z)$ of the SPCW is defined in Fig. \ref{bandfig}(b).}\label{GSfig2}
    \end{figure*}
  
  We apply a gradient descent algorithm for the SPCW geometry $n(\vec{r})$ (design variables) to minimize the objective function $F_{\text{total}}(n(\vec{r}))=F_c+F_{D2}+F_{t}$ with intermittent thermal excitations to avoid local extrema, as with simulated annealing. The objective function consists of the contributions from band curvature $F_c\propto |m_e|^{-2}$ and frequency deviations $F_{D2} (F_t)=|\omega_b-\nu_{D2}+\Delta_b|^2 (|\omega_b-\nu_{t}|^2)$ of $|F=4\rangle\rightarrow |F^{\prime}=5\rangle$ transition frequency $\nu_{D2}$ (blue-detuned magic wavelength frequency $\nu_t$) for atomic Cesium from the band edges $w_b$ of the respective modes. During the optimization sequence, the complex band diagram is computed to estimate the effective mass $m_e$ and the localization length $L_c$ with plane-wave expansions \cite{mpbcitation}. After convergence, we switch over to a finite structure with device length $L_d$ and apply a combination of filter-diagonalized FDTD and FDFD methods  \cite{Oskooi2010,Shin2012} on a high-bandwidth interconnected computational cluster with the Yee lattice modified to directly optimize the dyadic Green's function $\boldsymbol{G}(\boldsymbol{x},\boldsymbol{x}^\prime,\omega)$ \cite{Sakoda2005,Vlack2012} and arrive at the final design variable $n(\vec{r})$ in Table \ref{SPCWtable}. To include imperfections of realistic devices, we introduce the uncertainty $\pm 1$ nm to the system variables consistent with the state-of-art PCW nanofabrication techniques \cite{Yu2014}.
  
  \begin{table}[b]
  \caption{Final design variables for the SPCW with slab index $n=2$. The uncertainty $\pm 1$ nm is added for the normal distributions of the disordered SPCW structure in Fig. \ref{J1DGammafig}.} \label{SPCWtable}
  \footnotesize\begin{ruledtabular}
  \begin{tabular}{lc | lc}
  Lattice constant $a_0$ & $366\pm 1$ nm  & Slot width $w$ & $226\pm 1$ nm \\
  Slab thickness $t$ & $200\pm 1$ nm & Squircle radius $r_s$ & $99\pm 1$ nm \\
  Secondary radius $r^{\prime}$ & $105\pm 1$ nm & Hole radius $r$ & $109\pm 1$ nm \\
  First line shift $l$ & $413\pm 1$ nm & Secondary line shift $m$  & $729\pm 1$ nm \\
  Squircle height $a$ & $79\pm 1 $ nm & Squircle width $b$ & $124\pm 1 $ nm
  \end{tabular}
  \end{ruledtabular}
  \end{table}
  
  The result of dispersion engineering is shown in Fig. \ref{bandfig}(a) for our flat-band Silicon Nitride SPCW slab, with the effective mass $m_e=2.1$ $\text{Hz}^{-1}\cdot \text{m}^{-2}$. In the single-band approximation, the localization length is expected to be $L_c\simeq 2 a_0$ at $\Delta_e=0.4$ THz. We assume that the atom is confined by the blue-detuned magic-wavelength GM trap $\nu_{t}$ at $\lambda_t=793.5$ nm (blue line of Fig. \ref{bandfig}(a)) with the intensity represented by the blue-colored contour map in Fig. \ref{bandfig}(d). The excited states of the trapped atom is modified by the vacuum of $\nu_{D2}$-mode (red line of Fig. \ref{bandfig}(a)) as indicated by the red contour map in Fig. \ref{bandfig}(d). At the band edge $k_0=0.5$, $\nu_{D2}$-mode is highly localized with the effective mode area $A_{\text{eff}}\simeq 0.18\lambda_{D2}^2$. The resulting photonic lamb shift is $\Delta_{\text{1D}}\simeq 620$ MHz at $\Delta_e=0.4$ THz.

  We now turn to the numerical Green's function $\boldsymbol{G}(\boldsymbol{x},\boldsymbol{x}^\prime,\omega)$ of a finite SPCW with device length $L_d=80a_0$ in Fig. \ref{J1DGammafig}. We evaluate the collective decay and the coherent interaction
  \begin{eqnarray}
  \Gamma_{\text{total}}^{(i,j)}&=&\frac{\mu_0 \omega^2}{\hbar}\text{Im}[\boldsymbol{d}^{*}\cdot \boldsymbol{G}(\boldsymbol{x}_i,\boldsymbol{x}_j,\omega)\cdot\boldsymbol{d} ],\\
  \Delta_{\text{Lamb}}^{(i,j)}&=&\frac{2\mu_0 \omega^2}{\hbar}\text{Re}[\boldsymbol{d}^{*}\cdot \boldsymbol{G}_s(\boldsymbol{x}_i,\boldsymbol{x}_j,\omega)\cdot\boldsymbol{d} ],
  \end{eqnarray}
  where the scattering Green's function is $\boldsymbol{G}_s=\boldsymbol{G}-\boldsymbol{G}_0$ with respect to the vacuum $\boldsymbol{G}_0$. More generally, we also define the waveguide Green's function $\boldsymbol{G}_{wg}=\boldsymbol{G}-\boldsymbol{G}_h$ absent the homogeneous (non-guided) contributions $\boldsymbol{G}_h$ (coupling to the lossy modes beyond the light cone and to the free-space modes), where the waveguide portion $\boldsymbol{G}_{wg}$ can be estimated from a multimode cavity model \cite{Hughesstudentthesis} under a single-band approximation, with the resulting decay rate
  \begin{equation}
  \Gamma_{\text{1D}}=\frac{\mu_0 \omega^2}{\hbar}\text{Im}[\boldsymbol{d}^{*}\cdot \boldsymbol{G}_{wg}(\boldsymbol{x}_i,\boldsymbol{x}_j,\omega)\cdot\boldsymbol{d} ],
  \end{equation}
  into the waveguide GM.
  
  As shown in Fig. \ref{J1DGammafig}, in the dispersive regime  \cite{Goban2015}, the flat band $\nu_{D2}$ exhibits extreme slow-light enhancement of the decay rate with group index $n_g\simeq 1,000$ near the band edge. As the atom enters the band gap in the reactive regime $\Delta_e>0$    \cite{Hood2016}, the waveguide decay rate $\Gamma_{\text{1D}}$ from $\boldsymbol{G}_{wg}$ is exponentially suppressed (red dashed line in Fig. \ref{J1DGammafig}(b)), while the highly asymmetric Fano-like resonance of $\boldsymbol{G}_{wg}$ around the band edge gives rise to a photonic Lamb shift $\Delta_{\text{1D}}\simeq 620$ MHz (Fig. \ref{J1DGammafig}(a)) that greatly exceeds $\Gamma_{\text{total}}\simeq 60$ MHz ($\Gamma_{\text{1D}}\simeq 4$ kHz) in the band gap with figure of merit $\mathcal{F}> 10^4$ at $\Delta_e=0.4$ THz (Fig. \ref{J1DGammafig}(c)), indicating significant coherence fraction in the collective motion relative to the correlated phononic dissipation. With the close agreement between the numerical Green's function $\boldsymbol{G}$ (black lines) and the waveguide model $\boldsymbol{G}_{wg}$ (red dashed lines) in Fig. \ref{J1DGammafig}, we can reliably predict $\Gamma_{1D}$ from $\boldsymbol{G}_{wg}$ and the mechanical loss factor $\gamma_m$ from both $\boldsymbol{G}_{wg}$ and $\boldsymbol{G}$. Thanks to the large band flatness, we can operate as close as $\Delta_b=5$ THz ($\Delta_e\simeq 10$ THz) and attain short-ranged motional coupling over $L_c\sim 2 a_0\ll L_d$, while maintaining inherent figure of merit $\mathcal{F}\sim 10^{10}$. We remark that $\mathcal{F}$ is defined as the ultimate coherence-to-dissipation ratio for the collective phonon modes in Section \ref{Lamb_section}, where we only consider the inherent dissipation of the atomic motions in the photonic band gap. In practice, our method will be realistically limited by the phase-noises of Raman sideband lasers and the inhomogeneous hyperfine broadening of the trapped atoms, as well as various uncontrollable surface forces.
  
  \begin{figure*}[t!]
  \includegraphics[width=1.75\columnwidth]{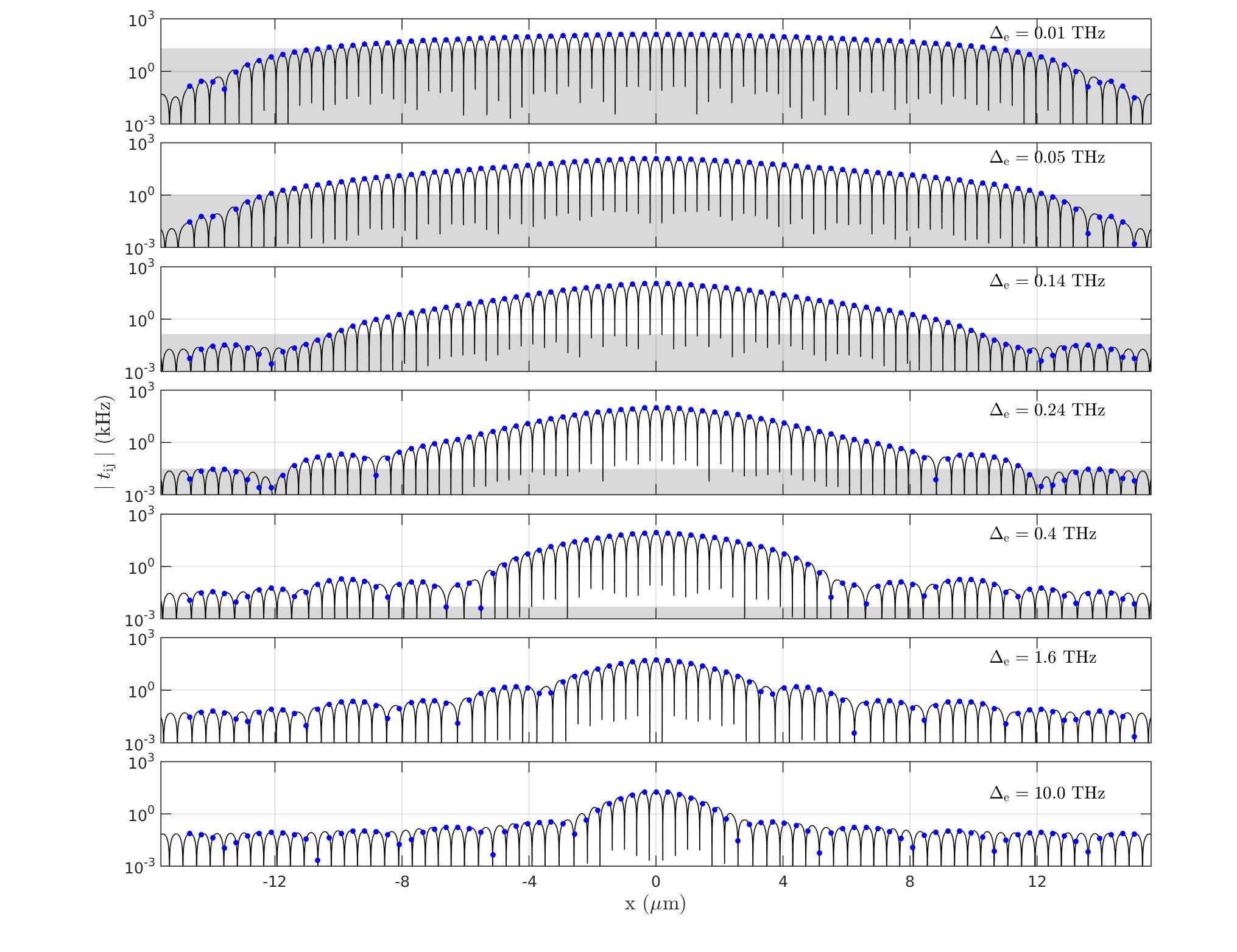}
  \caption{{Short-ranged atom-atom interaction in a photonic band gap.} We numerically evaluate the non-local Green's function $\boldsymbol{G}(\boldsymbol{x}_i,\boldsymbol{x}_j,w)$ for the SPCW and obtain the figure of merit for effective detunings $\Delta_e=0.01, 0.05, 0.14, 0.24, 0.4, 1.6, 10$ THz. Due to the large photon mass $m_e$, the atoms experience exponentially localized tunneling interactions $t_{ij}/\gamma_m\gg 1$ over lengths $L_c$. The grey shaded regions depict the dissipative regime with $t_{ij}<\gamma_m$, where collective phononic loss dominates over the coherent tunneling rate. For large $\Delta_e$, the ratio $t_{ij}/\gamma_m\gg 10^4$ is exponentially enhanced at the expense of reduced values $t_{ij}\simeq  2\pi\times 20$ kHz and localized length $L_c\simeq 2.5 a_0$ at $\Delta_e\simeq 10$ THz.}\label{NLoGFfig2}
  \end{figure*}
  
  For disordered photonic structures, we compute the dyadic Green's functions with the Gaussian-random geometric disorder $\sim 1$ nm (positions and sizes of the holes, thickness of the waveguide) distributed across the entire nanophotonic waveguide. In a single realization, the radiative enhancement factor at the band edge may be hindered by Anderson and weak localization. However, in the reactive regime $\Delta_e>0$, we observe that the decay rate and the photonic Lamb shift in Fig. \ref{J1DGammafig}, as well as the nonlocal Green's function $\boldsymbol{G}(\boldsymbol{x}_i,\boldsymbol{x}_j,\omega)$ are not significantly modified by the structural disorders $\sim 1$ nm (grey dashed lines in Fig. \ref{J1DGammafig}). Such nano-fabrication tolerances have been demonstrated in Refs.   \cite{Goban2014, Hood2016}. Because of the nature of the photonic bandgap, the non-radiative atom-field localized modes are resistant to the degree of structural disorder.
  
  \subsection{Ground-state potentials and phononic modes}
  We now turn our attention to the trapping mechanism for the atoms in the SPCW.
  To form an atomic chain, we confine the atoms in the $y-z$ plane by two incoherent side-illumination (SI) beams  \cite{Thompson2013, Goban2015} and localize the x-motion by a weak GM trap at $794$ nm, as shown in Fig. \ref{GSfig2}. With the SI beams near the blue-detuned magic wavelengths $\lambda=687$ nm in an optical accordion, we anticipate efficient loading into the GM trap. Because the SI beam provides additional confinement along $z$    \cite{Kanskar1997,Crozier2005,Bernard2016}, we can operate the GM trap away from the band edge at $k_x=0.48$, thereby reducing the intensity contrast along $x$. With this protocol we can gain a 3D FORT with trapping potential shown in Fig. \ref{GSfig2} (d--f).
  
  \begin{figure*}[th!]
    \includegraphics[width=1.5\columnwidth]{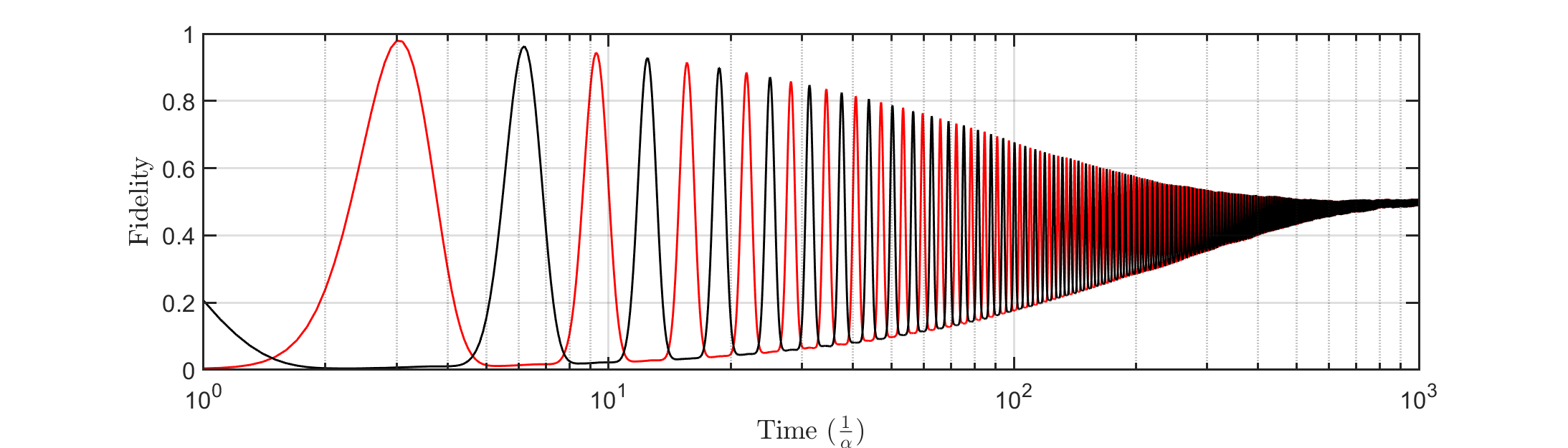}
    \caption{{Quantum-state transfer over a dissipative spin chain.} The open-system dynamics is numerically computed for the quantum-state transfer across $N=6$ atoms with figure of merit $\mathcal{F}\simeq 10^4$ by the quantum trajectory method. In addition to the intrinsic mechanical dissipation, we include spin-relaxation processes in the far-off-resonant optical trap. The state fidelity of the first (last) atom in the spin chain is displayed as a black (red) line.}\label{FigS8}
    \end{figure*}

  From the numerical non-local Green's function $\boldsymbol{G}(\boldsymbol{r}_i,\boldsymbol{r}_j,w)$, we observe that the localization length scales with $L_c=\sqrt{1/2m_e\Delta_e}$ and the effective mass $m_e=2.1$ $\text{Hz}^{-1}\cdot \text{m}^{-2}$ up to $\Delta_e\simeq 5$ THz. We attribute the deviation of the localization scaling beyond $\Delta_e>5$ THz to the residual Lamb shift by the off-resonant couplings to the other bands and to the slab modes. Fig. \ref{NLoGFfig2} depicts the local nature of external atom-atom interaction $t_{ij}= \eta_l^2 f^2 \Delta_{\text{Lamb}}(x_i,x_j)$  with $\eta_l=x_0/L_c$  relative to the mechanical decoherence $\gamma_m=\eta_l^2 f^2 (\Gamma_{\text{1D}}+|\Delta_{\text{Lamb}}/\Delta_e|^2\Gamma^{\prime})$, where the homogeneous decay rate $\Gamma^\prime\simeq 0.7 \Gamma_{\text{vac}}$ is weakly inhibited. At $\Delta_e=0.4$ THz, we find tunneling rate $t\simeq 2\pi\times 230$ kHz and localization length $L_c=0.77 \mu$m and phonon loss rate $\gamma_m\simeq 2\pi \times 5$ Hz. Another possible error source could be recoil heating from the trapping beam. Since we work with FORT in blue detuning, the heating rate can be estimated as $\gamma_\text{heat} \simeq E_r(\Omega_t/\delta_t)^2\Gamma^\prime/{\hbar w_t}$ \cite{Gerbier2010}, where $\Omega_t$ and $\delta_t$ are trapping Rabi frequency and laser-atom detuning respectively, and $E_r=4\pi^2\hbar^2/2m\lambda_t^2$ is recoil energy. For cesium atom and our trapping setup, the heating rate is estimated as $\gamma_\text{heat} \sim 0.2 \text{Hz} \ll \gamma_m$, therefore can be neglected safely.
  
  Beyond the scope of the present work, we have also investigated SPCWs with strong phononic on-site $U_0$ interactions, which maps the phononic model to XXZ spin magnet and Luttinger liquids for finite filling factor. Further design variation that provides strong phononic density-density interaction $U_{ij} \hat{n}_i \hat{n}_j$ will be discussed elsewhere. Such a constraint on the local phonon field provides a mechanism to impose local symmetry similar to the context of lattice gauge theories in condensed matter systems.
  
    \begin{table*}[t!]
  \begin{tabular}{|c|c|c|c||c|c|c|}
  \hline
  \tabincell{c}{Energy \\ Hierarchy}& Expression & Requirements & \tabincell{c}{Typical \\ Value} &\tabincell{c}{Effective \\ Error source} & Expression & \tabincell{c}{Typical \\ Value}  \\
  \hline\hline
  \tabincell{c}{Atom-PCW \\ interaction} & $g_c\simeq \sqrt{\frac{\omega_b d^2}{2\epsilon_0 A_{\text{eff}} L_c}}$ &  & $\sim10 \text{GHz}$ &\tabincell{c}{Photon \\ loss}&$\kappa\simeq\kappa_0 \exp(-L_d/L_c) +\frac{g_c^2}{\Delta_e^2}\Gamma^\prime$&$\sim 10 \text{MHz}$\\
  \hline
  \tabincell{c}{Mechanical \\ tunneling} & $ t_{ij}\simeq \eta_{l}^2 g_m$ & \tabincell{c}{$f \ll 1$}& $\sim 1 \text{MHz}$&\tabincell{c}{Phonon \\ loss}&$\gamma_m\simeq\frac{\eta_l^2g_m}{\Delta_e}\kappa$&$\sim 10 \text{Hz}$ \\
  \hline
  \tabincell{c}{Spin-spin \\ interaction} & $J^{(i,j)}_{\alpha, \beta}=2\text{Re}[\frac{{\tilde{\Omega}^{(i)}_{\alpha,l}\tilde{\Omega}^{(j)\ast}_{\beta,l}}}{\Delta_M}]$ & \tabincell{c}{$|\tilde{\Omega}^{(i)}_{\alpha,l}| \ll \Delta_M$ \\ $\ll |\epsilon_{l\pm1}-\epsilon_{l}|$} & $ \sim 50 \text{kHz}$ &\tabincell{c}{Spin \\ decoherence}&$\gamma^{(i,j)}_{\alpha,\beta}\simeq\frac{\gamma_m}{\Delta_M}J^{(i,j)}_{\alpha,\beta}$&$\sim 0.1 \text{Hz}$ \\
  \hline
  \end{tabular}
  \caption{A summary of energy scale hierarchy and corresponding effective error rates.}\label{table}
  \end{table*}

  \subsection{Phonon-mediated spin-exchange coefficient}
  For universal spin-control with $N\simeq 50$ atoms, we estimate the spin-exchange coupling rate $J_{ij}\simeq 50$ kHz with the intrinsic decoherence rate $\gamma_{\alpha,\beta}^{(i,j)}\ll  1$ Hz at $\Delta_e=0.4$ THz. As an example, we depict the open-system dynamics of the quantum-state transfer protocol in Fig. \ref{FigS8} by solving the master equation (Eqs. \ref{twobodyHam}-\ref{twobodyHamLindblad}). As discussed above, because of $\gamma_m/\Delta_M\sim 10^{-4}$, the intrinsic phonon-induced spin decoherence $\gamma_{\alpha,\beta}^{(i,j)}$ is highly negligible. We thereby include the spin-relaxation rate $\gamma^{(i,i)}_{\text{FORT}}<1$ Hz of the FORT beams  \cite{Cline1994} by adding the following local dissipative terms to the original master equation in Eq. \ref{twobodyHamLindblad}.
  \begin{eqnarray}
  \mathcal{L}_{ss} [\hat{\rho}_{\mathcal{S}}]&=&-\sum_i \frac{\gamma^{(i,i)}_{\text{FORT}}}{2}(\{\hat{\sigma}_{ss},\hat{\rho}_{\mathcal{S}}\}-2\hat{\sigma}_{gs}\hat{\rho}_{\mathcal{S}} \hat{\sigma}_{sg}),\nonumber\\
  \mathcal{L}_{gg} [\hat{\rho}_{\mathcal{S}}]&=&-\sum_i \frac{\gamma^{(i,i)}_{\text{FORT}}}{2}(\{\hat{\sigma}_{gg},\hat{\rho}_{\mathcal{S}}\}-2\hat{\sigma}_{sg}\hat{\rho}_{\mathcal{S}} \hat{\sigma}_{gs}).\nonumber
  \end{eqnarray}
  We note that, due to the highly differential decay rates for the $D1$ and $D2$ lines of Cs by the SPCW, we do not observe any suppression of Raman spontaneous emission rates relative to the Reyleigh scattering by the FORT. The state-fidelities for $N=1$ and $N=6$ atoms are displayed as black and red solid lines n Fig. \ref{FigS8}, respectively. We assume an initially injected spin state of $|s\rangle$ with the parameters of Fig. \ref{QSTfig}. For the clarity of presentation, the remaining spin-medium is prepared to the ground state $|g\cdots g\rangle$. As the spin-excitation is transferred within the dissipative spin chain, the overall spin medium is thermally depolarized by the actions of the local dissipation and the state-fidelity $F$ is progressively reduced to $F\rightarrow 0.5$ with $\hat{\rho}_{\mathcal{S}}\rightarrow \prod_i \frac{1}{2}(|g_i\rangle\langle g_i|+|s_i\rangle\langle s_i|)$.

\section{SU($N$)-GAUGED WAVEGUIDE QED SIMULATOR}

  \subsection{Generalized Gell-Mann matrices} \label{Appendix_Gell-mann}

  The $n$-dimensional Hermitian generalized Gell-Mann matrices (GGM) are the higher-dimensional extensions of the Pauli matrices (for qubit) and the Gell-Mann matrices (for qutrit). Similar to the roles which the Pauli (Gell-Mann) matrices play in SU($2$) (SU($3$)) algebra, they are the standard SU($n$) generators. There are three different types of GGMs --- $\frac{n(n-1)}{2}$ symmetric ones, $\frac{n(n-1)}{2}$ anti-symmetric ones and $n-1$ diagonal ones, which are defined respectively as
  \begin{enumerate}
  \item Symmetric GGMs ($1\le \alpha < \beta \le n$)
  \begin{equation}
  \hat{\Lambda}_{\alpha\beta}^{(s)}=|\alpha\rangle\langle \beta|+|\beta\rangle\langle \alpha|,
  \end{equation}
  \item Anti-symmetric GGMs ($1\le \alpha < \beta \le n$)
  \begin{equation}
  \hat{\Lambda}_{\alpha\beta}^{(a)}=-i|\alpha\rangle\langle \beta|+i|\beta\rangle\langle \alpha|,
  \end{equation}
  \item Diagonal GGMs ($1\le \alpha  \le n-1$)
  \begin{equation}
  \hat{\Lambda}_{\alpha\alpha}^{(d)}=\sqrt{\frac{2}{\alpha(\alpha+1)}}\sum_{\beta=1}^\alpha|\beta\rangle\langle \beta|-\alpha|\alpha+1\rangle\langle \alpha+1|,
  \end{equation}
  \end{enumerate}
  Hence, in total, we have $n^2-1$ GGMs. From the definitions, one can verify that, similar to the Pauli matrices, all GGMs are Hermitian and traceless. They are orthogonal and form a basis together with identity $\hat{I}_n$. 
  
  \subsection{Gauge-projected SU($n$) Heisenberg model}\label{Appendix_GPHeisenberg}
  
  To gauge the primitive Hamiltonian $\hat{H}_I$ to the local symmetry sector, we define a projection operator $\hat{P}_G$ which brings quantum states to the ground-state sector $\mathcal{Q}=n-2$ of the gauge Hamiltonian $\hat{H}_G$. Namely, $\hat{H}_G \hat{P}_G=\hat{P}_G \hat{H}_G=E_G \hat{H}_G$, where $E_G$ is the ground-state energy of $\hat{H}_G$. 
  
  We perturbatively expand $\hat{H}_I$ within the sector $\mathcal{Q}$ with the Kato's series
  \begin{eqnarray}
    \hat{H}_{\text{eff}}^{(1)}&=& \hat{P}_G \hat{H}_I \hat{P}_G\nonumber\\
    \hat{H}_{\text{eff}}^{(2)}&=& \hat{P}_G \hat{H}_I \hat{S}_{k1}\hat{H}_I \hat{P}_G\nonumber\\
    \hat{H}_{\text{eff}}^{(3)}&=& \hat{P}_G \hat{H}_I \hat{S}_{k1} \hat{H}_I \hat{S}_{k2} \hat{H}_I \hat{P}_G\nonumber\\
    &\cdots&,\nonumber
  \end{eqnarray}
  where $\hat{S}_0=-\hat{P}_G$ and $\hat{S}_n=[(1-\hat{P}_G)(E_G-\hat{H}_G)^{-1}]^n$. Because $\hat{H}_I$ breaks the local gauge symmetry, the first-order term vanishes $\hat{H}_{\text{eff}}^{(1)}=0$. The low-energy dynamics is thereby described at the second order with
  \begin{widetext}
  \begin{equation}
  \hat{H}_{\text{eff}}^{(2)}=\sum_{\overline{i},\overline{j}}  \mathcal{J}_{\overline{i},\overline{j}}\left(\sum_{\alpha\neq \beta}\hat{\sigma}_{+}^{(\alpha,\overline{i})}\hat{\sigma}_{-}^{(\beta,\overline{i})}\hat{\sigma}_{+}^{(\beta,\overline{j})}\hat{\sigma}_{-}^{(\alpha,\overline{j})}\prod_{k\neq \beta}\hat{\sigma}_{gg}^{(k,\overline{i})}\prod_{l\neq \alpha}\hat{\sigma}_{gg}^{(l,\overline{j})}+\hat{\sigma}_{ee}^{(\alpha,\overline{i})}\hat{\sigma}_{ee}^{(\beta,\overline{j})}\prod_{k\neq\alpha}\hat{\sigma}_{gg}^{(k,\overline{i})}\prod_{l\neq\beta}\hat{\sigma}_{gg}^{(l,\overline{j})}+h.c\right),\label{gauge-invariant-eq-appendix}
  \end{equation}
  \end{widetext}
  where the ring-exchange coefficient $\hat{J}_{\overline{i},\overline{j}}=-\mathcal{O}_{\overline{i},\overline{j}}^2/2\lambda_G$ is mediated by a pair of virtual spinon excitations $\mathcal{Q}^{\prime}=\mathcal{Q}\pm 2$. By the addition of a gauge-invariant $2$-body Hamiltonian $\hat{H}_{\text{anc}}=-\sum_{\overline{i},\overline{j}}\left (\mathcal{J}_{\overline{i},\overline{j}}\sum_{\alpha\neq \beta}\hat{\sigma}_{ee}^{(\alpha,\overline{i})}\hat{\sigma}_{ee}^{(\beta,\overline{j})}  + \mathcal{D}_{\overline{i},\overline{j}} \sum_{\alpha} \hat{\sigma}_{ee}^{(\alpha,\overline{i})}\hat{\sigma}_{ee}^{(\alpha,\overline{j})} \right)$ to the perturbative Hamiltonian $\hat{H}_{\text{total}}=\hat{H}_{\text{anc}}+\hat{H}_{\text{eff}}^{(2)}+\hat{H}_G$, we obtain the effective Hamiltonian in Eq. \ref{PhysicalSU3equation} within the single-excitation gauge sector $\mathcal{Q}$.

\subsection{Sachdev-Ye quantum magnet}\label{Appendix_SYModel}

\begin{figure}[t]
  \includegraphics[width=1\columnwidth]{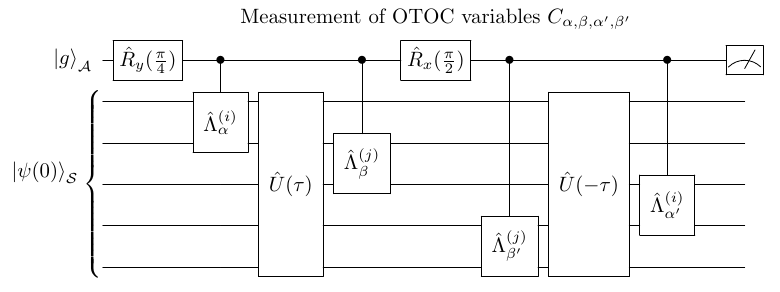}
  \caption{{Construction of SU($n$) OTOCs.} Measurement prescription of highly complex out-of-time-order correlators (OTOC). The circuit constructs the OTOC variables $C_{\alpha,\beta,\alpha^\prime,\beta^\prime}\equiv\langle\hat{\Lambda}^{(j)}_{\beta^\prime}(t)\hat{\Lambda}^{(i)}_{\alpha^\prime}(0) \hat{\Lambda}^{(j)}_\beta(t)\hat{\Lambda}^{(i)}_\alpha(0)\rangle$ of system atoms $\mathcal{S}$ and maps the values to the internal state of a single ancilla qubit $\mathcal{A}$. The time-inverse evolution for the global dynamics $\hat{U}(- \tau)=e^{- i(-\hat{H}_\text{SY})\tau}$ can be realized still in a positive time flow but with a negative Hamiltonian $-\hat{H}_\text{SY}$, i.e., inverting the sign of all $\mathcal{J}_{ij}$.}\label{fig5appendix}
  \end{figure}

In Section \ref{gaugedSUN_section}, we discussed the all-to-all connected SU($n$) Heisenberg model.  However, as an effective model in terms of second-order perturbation, all connections $\mathcal{J}_{ij}$ must be all negative (ferromagnetic) or positive (antiferromagnetic), determined by the eigenenergy sector that we choose, while fully Gaussian-random-distributed couplings are the crucial ingredients for the generation of quantum chaos of the Sachdev-Ye (SY) model  \cite{Sachdev1993, Sachdev2015}. SY Hamiltonian reads
\begin{equation}
\hat{H}_{\text{SY}}=\frac{1}{\sqrt{n}}\sum_{j>i}\mathcal{J}_{ij}\sum_{\alpha}\hat{\Lambda}_{\alpha}^{(i)}\hat{\Lambda}_{\alpha}^{(j)},
\end{equation}
where the after-quench connections $\{\mathcal{J}_{ij}\}$ are drawn from the probability distribution $P(\mathcal{J}_{ij}) \sim \exp{[-\mathcal{J}^2_{ij}/(2\mathcal{J}^2)]}$.

We describe a stroboscopic strategy to simulate the dynamics driven by such a Hamiltonian. For an arbitrary SY Hamiltonian, we can separate it into two parts $\hat{H}_\text{SY}=\hat{H}^{(+)}_\text{SY}+\hat{H}^{(-)}_\text{SY}$, where $\hat{H}^{(+)}_\text{SY}$ ($\hat{H}^{(-)}_\text{SY}$) contains only all terms with positive (negative) connections and thus can be realized efficiently in our platform. To realize a coarse-grained unitary evolution in a single time step $\Delta t$, we first turn on the positive Hamiltonian $\hat{H}^{(+)}_\text{SY}$ for a time period of $\Delta t/2$. Then, we switch on the $\hat{H}^{(-)}_\text{SY}$ for the same period and keep the Hamiltonian for another $\Delta t$. At last, we evolve the system again under $\hat{H}^{(+)}_\text{SY}$ for $\Delta t/2$. The entire dynamics is then given by $\exp{[{-i\hat{H}_\text{SY}\Delta t}]}+\mathcal{O}(\Delta t^3)$ with an error of the order $\Delta t^3$ due to the non-commuting $\hat{H}^{(+)}_\text{SY}$ and $\hat{H}^{(-)}_\text{SY}$.

We can also measure the out-of-time-operator-correlations for the SY model in our platform, which is essential for describing the entanglement scrambling in this system. The crucial step is creating a controlled GMM operation $\hat{U}_{C-\Lambda_\alpha}=|g\rangle\langle g| \otimes \hat{I} +|s\rangle\langle s| \otimes \hat{\Lambda}_\alpha $, which can be used to decompose an arbitrary SU($n$) operator. Let us take a controlled symmetric GGM C-$\hat{\Lambda}_{\alpha\beta}^{(s)}$ as an example. To realize this kind of controlled operations, we can couple an ancilla qubit to the $\alpha$th and the $\beta$th qubits in a single logical block with the two-body term $\hat{H}_{\alpha\beta}=\chi_\alpha\hat{\sigma}_{ss}^{(\mathcal{A})}\hat{\sigma}^{(\alpha)}_+ +\chi_\beta\hat{\sigma}_{ss}^{(\mathcal{A})}\hat{\sigma}^{(\beta)}_+ +h.c$. This leads to an effective interaction $\tilde{\chi}_{\alpha\beta}\hat{\sigma}_{ss}^{(\mathcal{A})}\hat{\mathcal{T}}_{\alpha\beta}+h.c.$ within the gauge-invariant sector $\mathcal{Q}$, where $\tilde{\chi}_{\alpha\beta}=\chi^\ast_\alpha\chi_\beta/\lambda_G$ and $\lambda_G$ is the coupling constant in gauge Hamiltonian $\hat{H}_G$ defined in Section \ref{gaugedSUN_section}. According to the definition of GGMs, if $\tilde{\chi}_{\alpha\beta}$ is real, the evolution under this Hamiltonian for an interaction time $t=\pi/{2|\tilde{\chi}_{\alpha\beta}|}$ yields $\hat{U}_{C-\Lambda^{(s)}_{\alpha\beta}}$.  And if we set $\tilde{\chi}_{\alpha\beta}$ as pure imaginary, a controlled anti-symmetric GGM C-$\hat{\Lambda}_{\alpha\beta}^{(a)}$ would be realized.  

We next describe a general method to construct and efficiently measure OTOCs for arbitrary SU($n$) observables in this system driven by arbitrary Hamiltonian $\hat{H}_\text{SY}$ without tomographic reconstruction. Unlike other protocols, our strategy is to encode the OTOC onto the single ancilla qubit $\mathcal{A}$ through controlled string operation and interferometrically read out the internal state of a \textit{single} ancilla qubit. We consider two operators, $\hat{V}^{(i)}=\sum_\alpha v^{(i)}_\alpha\hat{\Lambda}^{(i)}_\alpha$ and $\hat{W}^{(j)}=\sum_\beta w^{(j)}_\beta \hat{\Lambda}_\beta^{(j)}$, acting on the system logical magnons and decomposed by the GGM operators $\{\hat{\Lambda}^{(i)}_\alpha\}$ and $\{\hat{\Lambda}^{(i)}_\beta\}$. The goal is then to measure all $C_{\alpha,\beta,\alpha^\prime,\beta^\prime}\equiv\langle\hat{\Lambda}^{(j)}_{\beta^\prime}(\tau)\hat{\Lambda}^{(i)}_{\alpha^\prime}(0) \hat{\Lambda}^{(j)}_\beta(\tau)\hat{\Lambda}^{(i)}_\alpha(0)\rangle$ and construct the overall OTOC with weighted distribution $w^\ast_{\beta^\prime}v^\ast_{\alpha^\prime}w_\beta v_\alpha$. The circuit in Fig. \ref{fig5appendix}(b) facilitates the transformation that maps the dynamical correlators $C_{\alpha,\beta,\alpha^\prime,\beta^\prime}$ to the ancilla qubit with the initial system-ancilla state $|\psi(0)\rangle_{\mathcal{S}}\otimes|g\rangle_{\mathcal{A}}$. The ancilla atom can be physically represented by the atoms in close proximity to the impedance-matching tethers of PCWs, so that the internal spins of the ancilla atom can readily evanescently dissipate to the input and output couplers. The sequence of gate sets maps the initial state to $\hat{V}^{(i)}_{\alpha^\prime}(0)\hat{W}^{(j)}_{\beta^\prime}(\tau)|\psi(0)\rangle_\mathcal{S}|s\rangle_{\mathcal{A}}+\hat{W}^{(j)}_\beta(\tau)\hat{V}^{(i)}_\alpha(0)|\psi(0)\rangle_\mathcal{S}|g\rangle_{\mathcal{A}}$
 with $\hat{O}(\tau)=e^{i\hat{H}\tau}\hat{O}e^{-i\hat{H}\tau}$. Here, the time-inverse evolution can be realized in a positive time flow but inverse the sign of all $\mathcal{J}_{ij}$. As with Ramsey interferometer, we measure the expectation values of the local spin vectors for qubit $\mathcal{A}$, where the dynamic correlators of the system atoms are $C_{\alpha,\beta,\alpha^\prime,\beta^\prime}=\frac{1}{2}\left[ \langle\hat{\sigma}_x\rangle_{\mathcal{A}}+i\langle\hat{\sigma}_y\rangle_{\mathcal{A}}\right]$. This method can be extended to high-order dynamic correlations in a straightforward fashion.

\subsection{Wess-Zumino-Witten quantum field theory}\label{Appendix_WZW}
 
As a minimal SU($N$) model, we discuss the realization of a stringly conformal field theory with an integrable $1$D SU($3$) Heisenberg model. Here, we investigate the universal features of the SU($3$)$_1$ Wess-Zumino-Witten (WZW) quantum field theory of level $k=1$. In particular, we extract the conformal data by accessing the entanglement entropy for an $1$D SU($3$) Heisenberg model at the Ulmin-Lai-Sutherland (ULS) critical point, the parent Hamiltonian for generating a zoo of strongly-correlated ground states, such as those found in fractional quantum Hall systems. Because of the versatile programmability, the gauged waveguide QED simulator can be readily extended to the low-energy states of the $2$D anti-ferromagnetic SU($n$) model (Eq. \ref{PhysicalSU3equation}), which are described by the $(2+1)$D WZW conformal field theory and holographically connected to a $3$D Chern-Simons quantum gravity in the scaling limit.

\begin{figure}[t]
	\includegraphics[width=1\columnwidth]{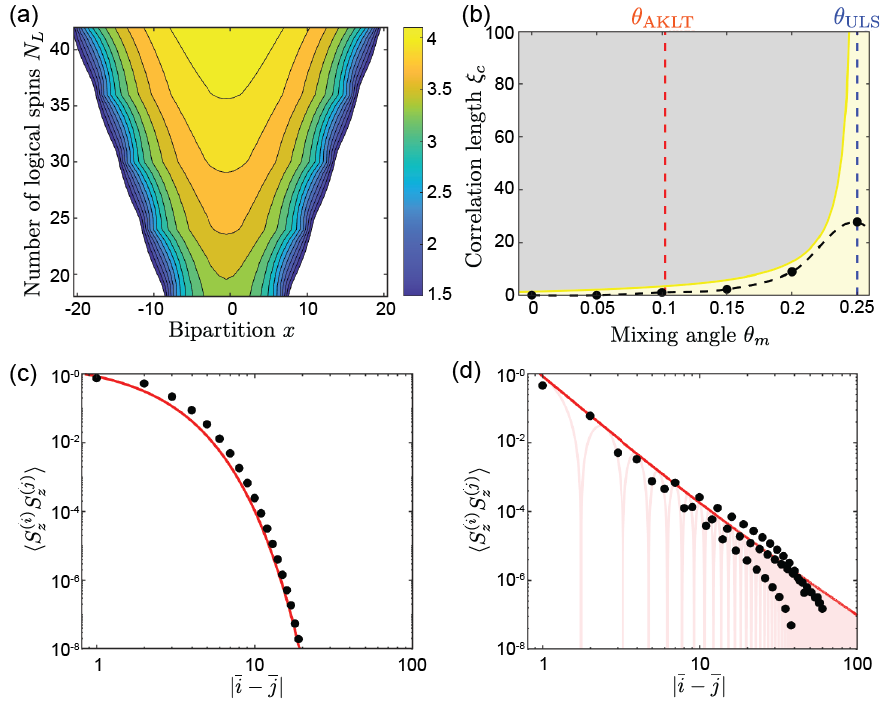}
	\caption{{Bilinear biquadratic spin-1 model.} (a) CFT scaling of entanglement entropy at the Uimin-Lai-Sutherland (ULS) point $\theta_{\text{ULS}}=\pi/4$. (b) Quantum phase transition between gapped Haldane phase and gapless nematic phase at the ULS quantum critical point. (c) Spin-spin correlation function $\langle \hat{S}_z^{(\overline{i})}\hat{S}_z^{(\overline{j})}\rangle$ at the  Affleck-Lieb-Kennedy-Tasaki (AKLT) point  $\theta_{\text{AKLT}}=\arctan(1/3)$ with a valence-bond ground state. (d) Spin-spin correlation function $\langle \hat{S}_z^{(\overline{i})}\hat{S}_z^{(\overline{j})}\rangle$ at the ULS point  $\theta_{\text{ULS}}=\pi/4$. The correlation functions and the phase diagram are computed from the uniform matrix product states (MPS), optimized by infinite DMRG algorithm with truncated bond dimension up to $\chi=500$. Finite $\chi$ generates an artificial cutoff in the correlation length $\xi_c$ to the otherwise algebraic correlation function. The fitting thereby only takes $|\overline{i}-\overline{j}|<\xi_c\simeq 40$ as the input. The entanglement entropy is simulated from a finite MPS for the logical SU($3$) spins ($3$ physical spins per logical spin) on a ring with bond dimension up to $\chi=8000$.}\label{WZW_appendix_fig_1}
\end{figure}

Namely, we consider the realization of a (1+1)D SU($3$)$_1$ WZW CFT for the antiferromagnetic SU($3$) Heisenberg model 
\begin{equation}
	\hat{H}_{\text{WZW}}=\mathcal{J}_c\sum_{\overline{i}}\sum_{\alpha} \hat{\Lambda}_{\alpha}^{(\overline{i})}\hat{\Lambda}_{\alpha}^{(\overline{i}+1)},\label{WZWHamiltonian}
\end{equation}
for the logical SU($3$) spins on a ring within the sector of $\mathcal{Q}=n-2$ of the waveguide QED simulator. Since $\mathcal{J}_c=-\mathcal{O}^2_{\overline{i},\overline{i}+1}/2\lambda_G<0$, the vacuum state of the WZW CFT is encoded onto the most excited state of Eq. \ref{WZWHamiltonian} within the sector $\mathcal{Q}$. This model has been extensively studied in the context of Haldane phase of the bilinear biquadratic (BBQ) spin-$1$ model
\begin{equation}
	\hat{{H}}_{\text{BBQ}}=\mathcal{J}_c\sum_{\overline{i}} \cos\theta \hat{\boldsymbol{S}}_{\overline{i}}\hat{\boldsymbol{S}}_{\overline{i}+1}+\sin\theta \left(\hat{\boldsymbol{S}}_{\overline{i}}\hat{\boldsymbol{S}}_{\overline{i}+1}\right)^2.\label{BBQHamiltonian}
\end{equation}
The enlarged SU($3$) symmetry of Eq. \ref{WZWHamiltonian} (Eq. \ref{BBQHamiltonian} at $\theta_{\text{ULS}}=\pi/4$) can be thought of as the consequence of the critical point of Berezinskii-Kosterlitz-Thouless (BKT) transition between the massive Haldane phase and an extended critical phase, described by the WZW field theory.

\subsubsection{Enlarged SU(3)-symmetry of bilinear biquadratic spin-1 models}
To understand the relationship between the familiar Haldane gap for spin-$1$ Heisenberg magnets at the exactly solvable point $\theta_{\text{AKLT}}=\arctan(1/3)$ (AKLT valence bond state) and the massless WZW field theory at $\theta_{\text{ULS}}=\pi/4$ (See also Fig. \ref{fig6}(b)), we describe how the SU($3$)-breaking marginal operator in the vicinity to the SU($3$)-symmetric critical point $\theta_{\text{ULS}}$ deforms the WZW CFT and dynamically generate a mass term in the Haldane phase by way of a BKT transition \cite{Sule2015}. By moving into the fermionic parton picture defined in Section \ref{WZW_section}, the BBQ Hamiltonian can be mapped to
\begin{equation}
	\hat{\mathcal{H}}_{\text{parton}}^{\text{BBQ}}= \hat{\mathcal{H}}_{\text{parton}}+\epsilon^2 \hat{\mathcal{H}}_{\text{marginal}},\label{Appendix_parton_eq}
\end{equation}
under the constraint $\sum_{\alpha}  \hat{\psi}^{(\overline{i})\dagger}_{\alpha}\hat{\psi}^{(\overline{i})}_{\alpha}=1$. The SU($3$)-symmetric parton Hamiltonian $\hat{\mathcal{H}}_{\text{parton}}$, defined in Eq. \ref{SUNsectionpartonEq}, is the dominant term near the ULS point, and kinetically exchanges excitations between the sites. The marginal operator $\hat{\mathcal{H}}_{\text{marginal}}=\mathcal{J}\sum_{\overline{i}} \hat{\psi}^{(\overline{i})\dagger}_{\alpha}\hat{\psi}^{(\overline{i})}_{\beta}\hat{\psi}^{(\overline{i}+1)\dagger}_{\alpha}\hat{\psi}^{(\overline{i}+1)}_{\beta}$, proportional to $\epsilon^2=\tan\theta-1$, projects the neighbouring sites to the singlet space, similar to the singlet projectors of the AKLT Hamiltonian. 

By applying the Hubbard-Stratotonovich transformation to Eq. \ref{Appendix_parton_eq} at the ULS point $\theta_{\text{ULS}}$, we obtain the mean-field Hamiltonian $\hat{H}_{\text{mf}}=|\chi_{\overline{i},\overline{i}+1}|^2 +\mu_{\overline{i}}( \hat{\psi}^{(\overline{i})}_{\alpha} \hat{\psi}^{(\overline{i})}_{\alpha}-1)- \chi_{\overline{i},\overline{i}+1} \hat{\psi}^{(\overline{i})\dagger}_{\alpha}\hat{\psi}^{(\overline{i}+1)}_{\alpha}+h.c$ with the auxiliary fields $\chi_{\overline{i},\overline{i}+1}=\langle \hat{\psi}^{(\overline{i})\dagger}_{\beta} \hat{\psi}^{(\overline{i}+1)}_{\beta}\rangle$ and the constraints expressed in terms of the chemical potential $\mu_{\overline{i}}$, with a Fermi sea filled up to the momentum $k_F=\pi/3$. Around this saddle point, the low-energy physics of Eq. \ref{Appendix_parton_eq} at $\theta_{\text{ULS}}$ is described by $\hat{\psi}^{(\overline{i})\dagger}_{\alpha}=e^{i k_F x_{\overline{i}}}\hat{\psi}_{L,\alpha}(x_{\overline{i}})+e^{-i k_F x_{\overline{i}}}\hat{\psi}_{R,\alpha}(x_{\overline{i}})$ with the chiral fermions $\hat{\psi}_{L,\alpha} (x), \hat{\psi}_{R,\alpha} (x)$ only populated  at the Fermi points, and write Eq. \ref{Appendix_parton_eq} as
\begin{eqnarray}
\hat{\mathcal{H}}(x)\simeq \pi v_F \int &dx& \sum_{\alpha,\beta} (\hat{j}_{R}^{\alpha,\beta}\hat{j}_{R}^{\beta,\alpha}+\hat{j}_{L}^{\alpha,\beta}\hat{j}_{L}^{\beta,\alpha})\nonumber\\
&&+\epsilon^2 (\hat{j}_{R}^{\alpha,\beta}\hat{j}_{R}^{\alpha,\beta}+\hat{j}_{L}^{\alpha,\beta}\hat{j}_{L}^{\alpha,\beta}),\label{Appendix_FT_EQ}
\end{eqnarray}
in terms of U($3$)-currents $\hat{j}_L^{\alpha,\beta} (\hat{j}_R^{\alpha,\beta})=\hat{\psi}_{L,\alpha }^{\dagger}\hat{\psi}_{L,\beta} (\hat{\psi}_{R,\alpha }^{\dagger}\hat{\psi}_{R,\beta})$ and Fermi velocity $v_F$.

\begin{figure}[t]
	\includegraphics[width=0.75\columnwidth]{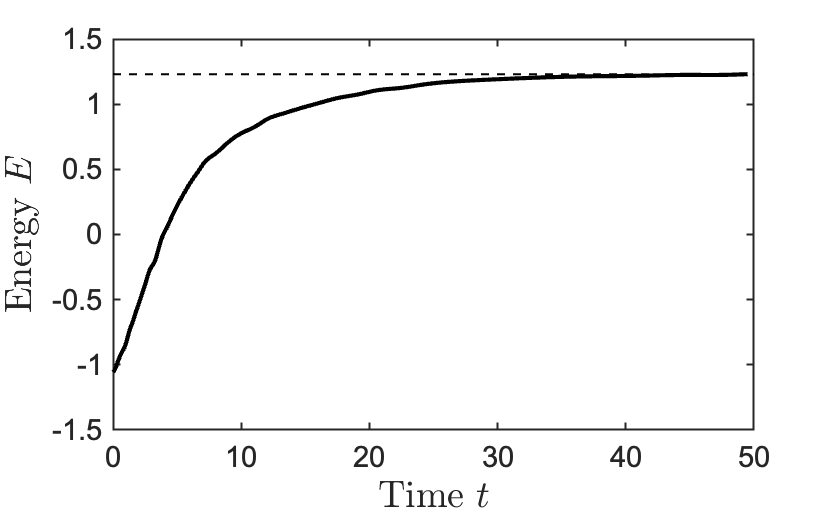}
	\caption{{Complex-time matrix-product state evolution.} Random matrix product state (MPS) is initially prepared for $54$ physical spins ($N_L=18$ logical spins), and the MPS is evolved under complex-time coordinate (Eq. \ref{WZW_appendix-complex-time-eq}) by way of time-evolving block decimation (TEBD) algorithm with an open boundary condition. At each time step, the SU($2$) MPS of the physical spins is transformed to the SU($3$)  MPS for the logical spins by locally contracting the SU($2$) MPS with an isometric matrix product operator (MPO) that projects the physical spins to the low-energy sector $\mathcal{Q}$. The overall dynamics is described by a cooling (heating) to (within) the ground state sector $\mathcal{Q}$, corresponding to the preparation of the vacuum state of the WZW CFT. The dashed line indicates the DMRG ground-state energy obtained for the target WZW Hamiltonian. The maximum bond dimension is $\chi=200$.}\label{WZW_appendix_fig_2}
\end{figure}

For the first term (with a global U($3$)$=$U($1$)$\oplus$SU($3$) symmetry), a U($1$) charge gap opens and leaves the SU($3$)-symmetric WZW model $\hat{\mathcal{H}}_{\text{WZW}}$ at the low-energy sector. Following the Abelian bosonization procedure $\hat{\psi}_{L,\alpha}= :1/2\pi\exp(-i\sqrt{4\pi}\hat{\phi}_{\alpha}):$ of Ref. \cite{Sule2015}, the SU($3$)-symmetric continuum Hamiltonian reads
\begin{equation}
\hat{\mathcal{H}}_{\text{WZW}}\sim \int dx ( \partial \phi_1 \partial \phi_1+\partial \phi_2 \partial \phi_2+\text{a.h})
\end{equation}
with two compact SU($3$) boson fields $\phi_{1,2}$ (each with central charge $c=1$). Fig. \ref{fig6}(a) depicts the CFT scaling behaviour of entanglement entropy following the Calabrese-Cardy formula for different system size $N_L$ ($c=2.05\pm 0.03$).  The entanglement entropy is computed by system-size expansion of finite matrix product states (MPS) for logical spins on a ring with a maximum bond dimension $\chi=8000$. The finite MPS was optimized using a hybrid complex-time evolution algorithm (Section \ref{Appendix_complex-time_section}). Following the operator product expansion, it can be shown that $\langle \hat{S}_z^{(\overline{i})} \hat{S}_z^{(\overline{j})}\rangle_{\text{ULS}}\sim \frac{\cos (2k_F |\overline{i}-\overline{j}|)}{|\overline{i}-\overline{j}|^{2\mathcal{D}}}$ with a scaling dimension $\mathcal{D}=2/3$ \cite{Itoi1997}. Fig. \ref{fig6}(d) displays the correlation function obtained by optimizing uniform MPS with infinite DMRG algorithm truncated to $\chi=500$, and the scaling dimension is fitted to $\mathcal{D}=0.68\pm0.03$.  The marginal perturbation of Eq. \ref{Appendix_FT_EQ}, on the other hand, breaks the global SU($3$) symmetry of the ULS point, and a mass gap $m_{\theta}=\exp[-\gamma (\theta_{\text{ULS}}-\theta)^{-0.6}]$ is dynamically generated for increasing coupling constant $\epsilon>0$ ($\theta<\theta_{\text{ULS}}$) with spin-spin correlation $\langle \hat{S}_z^{(\overline{i})} \hat{S}_z^{(\overline{j})}\rangle_{\theta}\sim \cos (2k_F |\overline{i}-\overline{j}|) e^{-m_{\theta} |\overline{i}-\overline{j}|}$ and non-universal constant $\gamma$. The asymptotic freedom of the marginal interaction at $\epsilon>0$ can be thought of as a BKT phase transition in terms of the renormalization group flow  \cite{Itoi1997}. The yellow line of Fig. \ref{fig6}(b) illustrates the scaling behavior of the correlation length $\eta_c\sim 1/m_{\theta}$ in comparison to those obtained from uniform MPS, where the non-universal constant $\gamma$ is fitted to the data points. The maximum correlation length $\xi_c\simeq 40$ at the ULS point is artificially cut off due to the finite $\chi=500$ truncation to the uniform MPS. 

\subsubsection{Hybrid complex-time algorithm}\label{Appendix_complex-time_section}

Because the vacuum state of WZW CFT corresponds to the most excited state within the low-energy sector $\mathcal{Q}$, standard DMRG algorithms cannot be adequately adapted to access the ground state of the target Hamiltonians (See also the inset of Fig. \ref{fig6}(a)). We instead apply a hybrid complex-time evolution to a random MPS in order to relax the system to the most excited state (target ground state) within the ground-state sector of the simulator by way of a time-evolving block decimation (TEBD) algorithm on the modified Hamiltonian
\begin{equation}
\hat{H}=\sum_{\overline{i},\overline{j}}(\hat{\mathcal{O}}_{\overline{i},\overline{j}}+\hat{D}_{\overline{i},\overline{j}})+i\hat{H}_G,\label{WZW_appendix-complex-time-eq}
\end{equation}
where the definitions of $\hat{\mathcal{O}}_{\overline{i},\overline{j}},\hat{D}_{\overline{i},\overline{j}},\hat{H}_G$ in Section \ref{gaugedSUN_section}. The imaginary constraint $\hat{H}_G$ allows the cooling of the random MPS to the sector $\mathcal{Q}$, while the first term mediates the gauge-invariant ring-exchange Hamiltonian (Eq. \ref{gauge-invariant-eq-appendix}) with an imaginary $\hat{J}_{\overline{i},\overline{j}}=i\mathcal{O}_{\overline{i},\overline{j}}^2/2\lambda_G$, which heats the system to the most excited state of the low-energy sector $\mathcal{Q}$. 

Following the complex-time evolution, an isometric matrix product projector (MPO) is locally contracted with the time-evolved MPS to map the physical SU($2$) spins to the logical SU($3$) spins. While the isometric tensor is not necessary to the protocol, we have found that such a practice allows a more intuitive interpretation on the operations taking place in the logical degrees of freedom. In particular, the converted MPS obtained through this method coincides to that obtained by performing a finite DMRG on the logical WZW Hamiltonian in Eq. \ref{WZWHamiltonian}. Fig. \ref{WZW_appendix_fig_2} displays the energy relaxation for the hybrid algorithm (solid line), which prepares the vacuum state of the WZW CFT (dashed line).

\bibliographystyle{apsrev4-2}

\begin{thebibliography}{1000}
\bibitem{Lloyd1996} S. Lloyd, Universal quantum simulators. \textit{Science} \textbf{273}, 1073 (1996).
\bibitem{Amico2008} L. Amico, R. Fazio, A. Osterloh, V. Vedral,  Entanglement in many-body systems. \textit{Rev. Mod. Phys.} \textbf{80}, 517 (2008).
\bibitem{Kimble2008} H. J. Kimble,  The quantum internet. \textit{Nature} \textbf{453}, 1023 (2008).
\bibitem{Bloch2008} I. Bloch, J. Dalibard, W. Zwerger, Many-body physics with ultracold gases. \textit{Rev. Mod. Phys.} \textbf{80}, 885 (2008).
\bibitem{Cubitt2018} T. S. Cubitt, A. Montanaro, S. Piddock, Universal quantum Hamiltonians. \textit{Proc. Nat. Acad. Sci.} \textbf{115}, 9497 (2018).
\bibitem{Biamonte2008} J. D. Biamonte, P. J.  Love,  Realizable Hamiltonians for universal adiabatic quantum computers. \textit{Phys. Rev. A} \textbf{78}, 012352 (2008).
\bibitem{Nagaj2008} D. Nagaj, P. Wocjan,  Hamiltonian quantum cellular automata in one dimension. \textit{Phys. Rev. A} \textbf{78}, 032311 (2008).
\bibitem{Vollbrecht2008} K. G. H. Vollbrecht, J. I. Cirac,  Quantum simulators, continuous-time automata, and translationally invariant systems. \textit{Phys. Rev. Lett.} \textbf{100}, 010501 (2008).
\bibitem{John1990} S. John, J. Wang,  Quantum electrodynamics near a photonic band gap: Photon bound states and dressed atoms. \textit{Phys. Rev. Lett.} \textbf{64}, 2418 (1990).
\bibitem{Kurizki1990} G. Kurizki,  Two-atom resonant radiative coupling in photonic band structures. \textit{Phys. Rev. A} \textbf{42}, 2915 (1990).
\bibitem{John1996} S. John, T. Quang, Quantum optical spin-glass state of impurity two-level atoms in a photonic bandgap. \textit{Phys. Rev. Lett.} \textbf{76}, 1320 (1996).
\bibitem{Hung2013} C.-L. Hung, S. M. Meenehan, D. E. Chang, O. Painter, H. J.  Kimble, Trapped atoms in one-dimensional photonic crystals. \textit{New. J. Phys.} \textbf{15}, 083026 (2013).
\bibitem{Thompson2013} J. D. Thompson, T. G. Tiecke, N. P. de Leon, J. Feist, A. V. Akimov, M. Gullans, A. S. Zibrov, V. Vuletic, M. D. Lukin, Coupling a single trapped atom to a nanoscale optical cavity. \textit{Science} \textbf{340}, 1202 (2013).
\bibitem{Tiecke2014} T. G. Tiecke, J. D. Thompson, N. P. de Leon, L. R. Liu, V. Vuletic, M. D. Lukin, Nanophotonic quantum phase switch with a single atom. \textit{Nature} \textbf{508}, 241 (2014).
\bibitem{Goban2014} A. Goban, C.-L. Hung, S.-P. Yu, J. D. Hood, J. A. Muniz, J. H. Lee, M. J. Martin, A. C. McClung, K. S. Choi, D. E. Chang, O. Painter, H. J. Kimble, Atom-light interactions in photonic crystals. \textit{Nature Comm.} \textbf{5}, 3808 (2014).
\bibitem{Goban2015} A. Goban, C.-L. Hung, J. D. Hood, S.-P. Yu, J. A. Muniz, O. Painter, H. J. Kimble, Superradiance for atoms trapped along a photonic crystal waveguide. \textit{Phys. Rev. Lett.} \textbf{114}, 063601 (2015).
\bibitem{Hood2016} J. D. Hood, A. Goban, A. Asenjo-Garcia, M. Lu, S.-P. Yu, D. E. Chang,  H. J. Kimble, Atom-atom interactions around the band edge of a photonic crystal waveguide. \textit{Proc. Natl. Sci. Am.} \textbf{113}, 10507 (2016).
\bibitem{Samutpraphoot2020} P. Samutpraphoot \textit{et al.} Strong coupling of two individually controlled atoms via a nanophotonic cavity. \textit{Phys. Rev. Lett.} \textbf{124}, 063602 (2020).
\bibitem{Dordevic2021} T. Dordevic \textit{et al.} Nanophotonic quantum interface and transportable entanglement for atom arrays. arXiv.2105.06485 (2021).
\bibitem{Douglas2015} J. S. Douglas,  H. Habibian, C.-L. Hung, A. V. Gorshkov, H. J. Kimble, D. E. Chang, Quantum many-body models with cold atoms coupled to photonic crystals. \textit{Nature Photon.} \textbf{9}, 326 (2015).
\bibitem{Shi2016} T. Shi, Y.-H. Wu, A. Gonzelez-Tudela, J. I. Cirac,  Bound states in boson impurity models. \textit{Phys. Rev. X} \textbf{6}, 021027 (2016).
\bibitem{Calajo2016} G. Calajo, F. Ciccarello, D. E. Chang, P. Rabl,  Atom-field dressed states in slow-light waveguide QED. \textit{Phys. Rev. A} \textbf{93}, 033833 (2016).
\bibitem{Gullans2012} M. Gullans, T. G. Tiecke, D. E. Chang, J. Feist, J. D. Thompson, J. I. Cirac, P. Zoller, M. D. Lukin, Nanoplasmonic lattices for ultracold atoms. \textit{Phys. Rev. Lett.} \textbf{109}, 235309 (2012).
\bibitem{Gonzalez-Tudela2015} A. Gonzalez-Tudela, C.-L. Hung, D. E. Chang, J. I. Cirac, H. J. Kimble, Subwavelength vacuum lattices and atom-atom interactions in two-dimensional photonic crystals. \textit{Nature Photon.} \textbf{9}, 320 (2015).
\bibitem{Hung2016} C.-L. Hung, A. Gonzelez-Tudela, J. I. Cirac, H. J. Kimble,  Quantum spin dynamics with pairwise-tunable, long-range interactions. \textit{Proc. Natl. Acad. Sci.} \textbf{113}, E4946 (2016).
\bibitem{Hartmann2006} M. J. Harmann, F. G. S. L. Brandao, M. B. Plenio,  Strongly interacting polaritons in coupled arrays of cavities. \textit{Nature Phys.} \textbf{2}, 849 (2006).
\bibitem{Greentree2006} A. D. Greentree, C. Tahan, J. H. Cole, L. C. L. Hollenberg,  Quantum phase transitions of light. \textit{Nature Phys.} \textbf{2}, 856 (2006).
\bibitem{Ramos2014} T. Ramos, H. Pichler, A. J. Daley, P. Zoller, Quantum spin dimers from chiral dissipation in cold-atom chains. \textit{Phys. Rev. Lett.} \textbf{113}, 237203 (2014).
\bibitem{Lodahl2017} P. Lodahl, S. Mahmoodian, S. Stobbe, A. Rauschenbeutel, P. Schneeweiss, J. Volz, H. Pichler, P. Zoller, Chiral quantum optics. \textit{Nature} \textbf{541}, 473 (2017).
\bibitem{Pichler2017} H. Pichler, S. Choi, P. Zoller,  M. D. Lukin, Universal photonic quantum computation via time-delayed feedback. \textit{Proc. Acad. Sci.} \textbf{114}, 11362 (2017).
\bibitem{Manzoni2017} M. T. Manzoni, L. Mathey, D. E. Chang, Designing exotic many-body states of atomic spin and motion in photonic crystals. \textit{Nature Comm.} \textbf{8} 14696 (2017).
\bibitem{Balents2010} L. Balents, Spin liquids in frustrated magnets. \textit{Nature} \textbf{464}, 199 (2010).
\bibitem{Sachdev2015} S. Sachdev, Bekenstein-Hawking entropy and strange metals. \textit{Phys. Rev. X} \textbf{5} 041025 (2015).
\bibitem{Kitaev2015} A. Y. Kitaev,  A simple model of quantum holography. \textit{Talks at KITP program: Entanglement in Strongly-Correlated Quantum Matter.} (2015).
\bibitem{Witten1984} E. Witten, Non-abelian bosonization in two dimensions. \textit{Commun. Math. Phys.} \textbf{92}, 455 (1984).
\bibitem{John1987} S. John, Strong localization of photons in certain disordered dielectric superlattices. \textit{Phys. Rev. Lett.} \textbf{58}, 2486 (1987).
\bibitem{Gruner1996} T. Gruner, D.-G. Welsch, Green function approach to the radiation-field quantization for homogeneous and inhomogeneous Kramers-Kronig dielectrics. \textit{Phys. Rev. A} \textbf{53}, 1818 (1996).
\bibitem{Dung2002}Ho Trung Dung, Ludwig Kn\"{o}ll and Dirk-Gunnar Welsch. Resonat dipole-dipole interaction in the presence of dispersing and absorbing surroundings. \textit{Phys. Rev. A} \textbf{66}, 063810 (2002).
\bibitem{Hughesstudentthesis}  G. Angelatos, \textit{Theory and applications of light-matter interactions in quantum dot nanowire photonic crystal systems.} PhD. Thesis (Queens University, Kingston, Canada, 2015).
\bibitem{Asenjo-Garcia2017} A. Asenjo-Garcia, J. D. Hood, D. E. Chang, H. J. Kimble,  Atom-light interactions in quasi-one-dimensional nanostructures: A Green's-function perspective. \textit{Phys. Rev. A} \textbf{95}, 033818 (2017).
\bibitem{Asenjo-Garcia2017b} A. Asenjo-Garcia, M. Moreno-Cardoner, A. Albrecht, H. J. Kimble, D. E. Chang, Exponential improvement in photon storage fidelities using subradiance and selective radiance in atomic arrays. \textit{Phys. Rev. X} \textbf{7}, 031024 (2017).
\bibitem{Reiter2012} Florentin. Reiter and Anders S. S\o rensen, Effective operator formalism for open quantum systems. \textit{Phys. Rev. A} \textbf{85}, 032111 (2012).
\bibitem{Jaksch2003} D. Jaksch, P. Zoller,  Creation of effective magnetic fields in optical lattices: The Hofstadter butterfly for cold neutral atoms. \textit{New J. Phys.} \textbf{5}, 56 (2003).
\bibitem{Bermudez2011} A. Bermudez, T. Schaetz, D. Porras, Synthetic gauge fields for vibrational excitations of trapped ions. \textit{Phys. Rev. Lett.}, \textbf{107}, 150501 (2011).
\bibitem{Korenbilt2012} S. Korenblit, D. Kafri, W. C. Campbell, R. Islam, E. E. Edwards, Z.-X. Gong, G.-D. Lin, L.-M. Duan, J. Kim, K. Kim, C. Monroe, Quantum simulation of spin models on an arbitrary lattice with trapped ions. \textit{New J. Phys.} \textbf{14}, 095024 (2012).
\bibitem{Aidelsburger2013}  M. Aidelsburger, M. Atala, M. Lohse, J. T. Barreiro, B. Paredes,  I. Bloch, Realization of the Hofstadter Hamiltonian with ultracold atoms in optical lattices. \textit{Phys. Rev. Lett.} \textbf{111}, 185301 (2013).
\bibitem{Bose2003} S. Bose, Quantum communication through an unmodulated spin chain. \textit{Phys. Rev. Lett.} \textbf{91}, 207901 (2003).
\bibitem{Christandl2004} M. Christandl, N. Datta, A. Ekert, A. J. Landahl, Perfect state transfer in quantum spin networks. \textit{Phys. Rev. Lett.} \textbf{92}, 187902 (2004).
\bibitem{Paternostro2005}  M. Paternostro, G. M. Palma, M. S. Kim, C. Falci, Quantum-state transfer in imperfect artificial spin networks. \textit{Phys. Rev. A} \textbf{71}, 042311 (2005).
\bibitem{Yung2005} M.-H. Yung, S. Bose, Perfect state transfer, effective gates, and entanglement generation in engineered bosonic and fermionic networks. \textit{Phys. Rev. A} \textbf{71}, 032310 (2005).
\bibitem{DiFranco2008} C. Di Franco, M. Paternostro, M. S. Kim, Perfect state transfer on a spin chain without state initialization. \textit{Phys. Rev. Lett.} \textbf{101}, 230502 (2008).
\bibitem{Yao2011} N. Y. Yao, L. Jiang, A. V. Gorshkov, P. C. Maurer, G. Giedke, J. I. Cirac, M. D. Lukin, Scalable architecture for a room temperature solid-state quantum information processor. \textit{Nature Comm.} \textbf{3}, 800 (2011).
\bibitem{Yao2013} N. Y. Yao, C. R. Laumann, A. V. Gorshkov, H. Weimer, L. Jiang, J. I. Cirac, P. Zoller, M. D. Lukin, Topologically protected quantum state transfer in a chiral spin liquid. \textit{Nature Comm.} \textbf{4}, 1585 (2013).
\bibitem{Kalmeyer1987} V. Kalmeyer, R. B. Laughlin,  Equivalence of the resonating-valence-bond and fractional quantum Hall states. \textit{Phys. Rev. Lett.} \textbf{59}, 2095 (1987).
\bibitem{Bauer2014} B. Bauer, L. Cincio, B. P. Keller, M. Dolfi, G. Vidal, S. Trebst, A. W. W. Ludwig, Chiral spin liquid and emergent anyons in a Kagome lattice Mott insulator. \textit{Nature Comm.} \textbf{5}, 5137 (2014).
\bibitem{Kumar2015} K. Kumar, K. Sun, E. Fradkin,  Chiral spin liquids on the Kagome Lattice. \textit{Phys. Rev. B} \textbf{92}, 094433 (2015).
\bibitem{Essafi2015} K. Essafi, O. Benton, L. D. C. Jaubert,  A Kagome map of spin liquids from XXZ to Dzyaloshinskii-Moriya ferromagnet. \textit{Nature Comm.} \textbf{7}, 10297 (2015).
\bibitem{Balents2002} L. Balents, M. P. A. Fisher, S. M. Girvin,  Fractionalization in an easy-axis Kagome antiferromagnet. \textit{Phys. Rev. B} \textbf{65}, 224412 (2002).
\bibitem{Isakov2011} S. V. Isakov, M. B. Hastings, R. G. Melko,  \textit{Nature Phys.} Topological entanglement entropy of a Bose-Hubbard spin liquid. \textbf{7}, 772 (2011).
\bibitem{Gingras2014} M. J. P. Gingras, P. A.  McClarty,  Quantum spin ice: a search for gapless quantum spin liquids in pyrochlore magnets. \textit{Rep. Prog. Phys.} \textbf{77}, 056501 (2014).
\bibitem{Castelnovo2008} C. Castelnovo, R. Moessner, S. L. Sondhi,  Magnetic monopoles in spin ice. \textit{Nature} \textbf{451}, 42 (2008).
\bibitem{Gorshkov2010} A. V. Gorshkov, M. Hermele, V. Gurarie, C. Xu, P. S. Julienne, J. Ye, P. Zoller, E. Demler, M. D. Lukin, A. M. Rey, Two-orbital SU(N) magnetism with ultracold alkaline-earth atoms. \textit{Nature Phys.} \textbf{6}, 289 (2010).
\bibitem{Tokura2000} Y. Tokura, N. Nagaosa, Orbital physics in transition-metal oxides. \textit{Science} \textbf{288}, 462 (2000).
\bibitem{Read1989} N. Read, S. Sachdev, Valence-bond and spin-Peierls ground states of low-dimensional quantum antiferromagnets. \textit{Phys. Rev. Lett.}, \textbf{62}, 1694 (1989).
\bibitem{Marston1989} J. B. Marston, I. Affleck,  Large-n limit of the Hubbard-Heisenberg model. \textit{Phys. Rev. B} \textbf{39}, 11538 (1989).
\bibitem{Greiter2007} M. Greiter, S. Rachel, Valence bond solids for SU(n) spin chains: Exact models, spinon confinement, and the Haldane gap. \textit{Phys. Rev. B} \textbf{75}, 184441 (2007).
\bibitem{Affleck1989} I. Affleck, Quantum spin chains and the Handane gap. \textit{J. Phys.: Cond. Matt.} \textbf{1}, 3047 (1989).
\bibitem{Sachdev1993} S. Sachdev, J. Ye,  Gapless spin-fluid ground state in a random quantum Heisenberg magnet. \textit{Phys. Rev. Lett.} \textbf{70}, 3339 (1993).
\bibitem{Kitaev2014} A. Kitaev,  Hidden correlations in the hawking radiation and thermal noise. \textit{Talks at the Fundamental Physics Prize Symposium.} (2014).
\bibitem{Shenker2014} S. H. Shenker, D. Stanford,  Black holes and the butterfly effect. \textit{J. High Energy Phys.} \textbf{3}, 67 (2014).
\bibitem{Maldecena2016} J. Maldacena, S. H. Shenker, D. Stanford, A bound on chaos. \textit{J. High Energy Phys.} \textbf{8}, 106 (2014).
\bibitem{Wess1971} J.  Wess, B.  Zumino, Consequences of anomalous ward identities. \textit{Phys. Lett. B} \textbf{37}, 95 (1971).
\bibitem{Witten1983} E. Witten, Global aspects of current algebra. \textit{Nuc. Phys. B} \textbf{223}, 422 (1983).
\bibitem{Witten1989} E. Witten, Quantum field theory and the Jones polynomial. \textit{Comm. Phys.} \textbf{121}, 351 (1989).
\bibitem{Sule2015} O. M. Sule, H. J. Changlani, I. Maruyama, S. Ryu. Determination of Tomonaga-Luttinger parameters for a two-component liquid. \textit{Phys. Rev. B} \textbf{92}, 075128 (2015).
\bibitem{Holzhey1994} C. Holzhey, F. Larsen, F. Wilczek. Geometric and renormalized entropy in conformal field theory. \textit{Nuc. Phys. B}, \textbf{424}, 443 (1994).
\bibitem{Calabrese2009} P. Calabrese, J. Cardy. Entanglement entropy and conformal field theory. \textit{J. Phys. A: Math. and Theo.} \textbf{42}, 504005 (2009).
\bibitem{Dalmonte2018} M. Dalmonte, B. Vermersh, P. Zoller. Quantum simulation and spectroscopy of entanglement Hamiltonians. \textit{Nature Phys.} \textbf{14}, 827 (2018).
\bibitem{Li2008} H. Li, F. D. M. Haldane, Entanglement spectrum as a generalization of entanglement entropy: Identification of topological order in non-Abelian fraction quantum Hall effect states. \textit{Phys. Rev. Lett.} \textbf{101}, 010504 (2008).
\bibitem{Senko2014} C. Senko, J. Smith, P. Richerme, A. Lee, W. C. Campbell, C. Monroe. Coherent imaging spectroscopy of a quantum many-body spin system. \textit{Science} \textbf{345}, 430 (2014).
\bibitem{Barredo2016} D. Barredo, S.  de Leseleuc, V. Lienhard, T. Lahaye, A. Browaeys, An atom-by-atom assembler of defect-free arbitrary 2D atomic arrays. \textit{Science} \textbf{354}, 1021 (2016).
\bibitem{Endres2016} M. Endres, H. Bernien, A. Keesling, H. Levine, E. R. Anschuetz, A. Krajenbrink, C. Senko, V. Vuletic, M. Greiner, M. D. Lukin, Atom-by-atom assembly of defect-free one-dimensional cold atom arrays. \textit{Science} \textbf{354}, 1024 (2016).
\bibitem{Kim2016} H. Kim, W. Lee, H.-G. Lee, H. Jo, Y. Song, J. Ahn, In situ single-atom array synthesis using dynamic holographic optical tweezers. \textit{Nature Comm.} \textbf{7}, 13317 (2016).
\bibitem{Meng2017} Y. Meng, A. Dareau, P. Schneeweiss, A. Rauschenbeutel, Near-ground-state cooling of atoms optically trapped 300 nm away from a hot surface. arXiv.1712.05749 (2017).
\bibitem{McMahon2016} P. L. McMahon, A. Marandi, Y. Haribara, R. Hamerly, C. Langrock, S. Tamate, T. Inagaki, H. Takesue, S. Utsunomiya, K. Aihara, R. L. Byer, M. M. Fejer, H. Mabuchi, Y. Yamamoto, A fully-programmable 100-spin coherent Ising machine with all-to-all connections. \textit{Science} \textbf{354}, 614 (2016).
\bibitem{Haribara2017} Y. Haribara, H. Ishikawa, S. Utsunomiya, K. Aihara, Y. Yamamoto, Performance evaluation of coherent Ising machines against classical neural networks. \textit{Quantum Sci. Technol.} \textbf{2}, 044002 (2017).
\bibitem{Poggi2016} P. M. Poggi, D. A. Wisniacki, Optimal control of many-body quantum dynamics: Chaos and complexity. \textit{Phys. Rev. A} \textbf{94}, 033406 (2016).
\bibitem{Yu2014} S.-P. Yu \textit{et al.} Nanowire photonic crystal waveguides for single atom trapping and strong light-matter interactions. \textit{Appl. Phys. Lett.} \textbf{104}, 111103 (2014).
\bibitem{Krauss2007}  Krauss, T. F.  {Slow-light in photonic crystal waveguides}. \textit{J. Phys. D: App. Phys.} \textbf{40}, 2666 (2007).
\bibitem{Arcari2014} Arcari, M. \textit{et al.} {Near-unity coupling efficiency of a quantum emitter to a photonic crystal waveguide}, \textit{Phys. Rev. Lett.} \textbf{113}, 093603 (2014).
\bibitem{Lodahl2015} Lodahl, P., Mahmoodian, S. \& Stobbe, S. {Interfacing single photons and single quantum dots with photonic nanostructures}, \textit{Rev. Mod. Phys.} \textbf{87}, 347 (2015).
\bibitem{mpbcitation}  Johnson, S. G. \&  Joannopoulos, J. D. {Block-iterative frequency-domain methods for Maxwell's equations in a plane-wave basis}, \textit{Opt. Exp.} \textbf{8}, 173 (2001).
\bibitem{Oskooi2010} Oskooi, A. F. \textit{et al.} {MEEP: A flexible free-software package for electromagnetic simulations by the FDTD method}, \textit{Comp. Phys. Comm.} \textbf{181}, 687 (2010).
\bibitem{Shin2012} Shin, W. \& Fan, S. {Choice of the perfectly matched layer boundary condition for frequency-domain Maxwell's equations solvers}, \textit{J. Comp. Phys.} \textbf{231}, 3406 (2012).
\bibitem{Sakoda2005} Sakoda, K. \textit{Optical properties of photonic crystals}, (Springer, 2005).
\bibitem{Vlack2012} Van Vlack, C. P. \& Hughes, S.  {Finite-difference time-domain technique as an efficient tool for calculating the regularized Green function: applications to the local-field problem in quantum optics for inhomogeneous lossy materials}, \textit{Opt. Lett.} \textbf{37}, 2880 (2012).
\bibitem{Kanskar1997} Kanskar, M. \textit{et al.} {Observation of leaky slab modes in an air-bridged semiconductor waveguide with a two-dimensional photonic lattice}, \textit{App. Phys. Lett.} \textbf{70}, 1438 (1997).
\bibitem{Crozier2005} Crozier, K. B. \textit{et al.} {Air-bridged photonic crystal slabs at visible and near-infrared wavelengths}, \textit{Phys. Rev. B} \textbf{73}, 115126 (2005).
\bibitem{Bernard2016} Bernard, S. \textit{et al.} {Precision resonance tuning and design of SiN photonic crystal reflectors}, \textit{Opt. Lett.} \textbf{41}, 5624 (2016).
\bibitem{Gerbier2010} Gerbier, F. \&  Castin, Y. {Heating rates for an atom in a far-detuned optical lattice}, \textit{Phys. Rev. A} \textbf{82}, 013615 (2010).
\bibitem{Cline1994}  Cline, R. A. \textit{et al.} {Spin relaxation of optically trapped atoms by light scattering}, \textit{Opt. Lett.} \textbf{19}, 207 (1994).
\bibitem{Itoi1997} C. Itoi, M.-H. Kato, Extended massless phase and the Haldane phase in a spin-1 isotropic antiferromagnetic chain. \textit{Phys. Rev. A} \textbf{55}, 8295 (1997).
\end{thebibliography}

\end{document}